\documentclass [prl,12pt,superscriptaddress]{revtex4}
\usepackage{graphicx}  
\usepackage{subfigure}
\usepackage{float}
\usepackage{dcolumn}   
\usepackage{bm}        
\usepackage{amsmath}        
\usepackage{amssymb}   
\usepackage{natbib}
    \DeclareMathOperator{\sech}{sech}
\begin{document}
\title{Microscopic theory for electrocaloric effects in planar double layer systems}

\author{Rajeev Kumar}
\email{kumarr@ornl.gov}
\affiliation{Center for Nanophase Materials Sciences, Oak Ridge National Laboratory, Oak Ridge, TN-37831}
\affiliation{Computer Science and Mathematics Division, Oak Ridge National Laboratory, Oak Ridge, TN-37831}

\author{Jyoti P. Mahalik}
\affiliation{Computer Science and Mathematics Division, Oak Ridge National Laboratory, Oak Ridge, TN-37831}

\author{Evgheni Strelcov}
\affiliation{Center for Nanophase Materials Sciences, Oak Ridge National Laboratory, Oak Ridge, TN-37831}

\author{Alexander Tselev}
\affiliation{Center for Nanophase Materials Sciences, Oak Ridge National Laboratory, Oak Ridge, TN-37831}

\author{Bradley S. Lokitz}
\affiliation{Center for Nanophase Materials Sciences, Oak Ridge National Laboratory, Oak Ridge, TN-37831}

\author{Sergei.V. Kalinin}
\affiliation{Center for Nanophase Materials Sciences, Oak Ridge National Laboratory, Oak Ridge, TN-37831}

\author{Bobby G. Sumpter}
\affiliation{Center for Nanophase Materials Sciences, Oak Ridge National Laboratory, Oak Ridge, TN-37831}
\affiliation{Computer Science and Mathematics Division, Oak Ridge National Laboratory, Oak Ridge, TN-37831}

\date{\today}

\begin{abstract}

\noindent We present a field theory approach to study changes in local temperature due to an applied 
electric field (the electrocaloric effect) in electrolyte solutions. Steric effects 
and a field-dependent 
dielectric function are found to be of paramount importance for accurate estimations of the 
electrocaloric effect. Interestingly, electrolyte solutions are found to exhibit negative electrocaloric effects. 
Overall, our results point toward using fluids near room temperature with low heat capacity and high salt concentration 
for enhanced electrocalorics.
\end{abstract}

\maketitle

There has been a renewed interest in
developing caloric materials\cite{caloricbook,caloricNMAT,caloricREV,landaubook} and advancing 
technologies\cite{caloricbook,caloricREV} 
for various refrigeration applications.
The caloric materials undergo reversible thermal changes under the influence of an applied
field, which can be magnetic, electric or mechanical in nature. These thermal changes due to magnetic 
field, electric field
and mechanical stresses are known as the magnetocaloric, electrocaloric and mechanocaloric effects, respectively.
Thermodynamic description of these changes was provided by Thomson\cite{thomson} and
the changes are results of variations in entropy of the system under the influence of an applied field.
The magnetocaloric effect is already used to reach temperatures in the milliKelvin (mK) 
range and is in the stage of being
commercialized for household refrigeration. In contrast, search for novel materials that can achieve the so-called
colossal or giant electrocaloric effect is currently a topic of extensive research. Historically, ferroelectric
materials\cite{caloricbook,caloricNMAT,caloricREV,landaubook,linesbook}, which are crystals with net 
polarization in the absence of any external applied electric field,
have been studied extensively for the electrocaloric effect and have shown thermal changes as low as $0.003$ K
near room temperature and as high as $31$ K based on the operating temperature and applied electric field\cite{caloricREV}.
It is to be noted that most of the materials studied for the electrocaloric effects are in the solid state
except some polymeric films\cite{caloricbook,caloricNMAT,caloricREV,neese}, which are considered 
viscoelastic, and liquid crystalline fluids\cite{liquidECE} that have shown giant electrocaloric effects in thin film 
geometries. 

Thermodynamic description of the electrocaloric effect\cite{linesbook} in an adiabatic system relies on the fact that
changes in entropy resulting from application of an electric field must be zero. As entropy
can be modified by varying either temperature ($T$) or difference in the surface potentials of the 
electrodes ($V_s$),
we consider infinitesimal changes in entropy ($\Delta S$) for a system at an initial temperature of $T = T_0$ and 
the potential difference
$V_s = V_0$ undergoing infinitesimal changes in the temperature ($\Delta T$) and the potential difference ($\Delta V_s$) so that
\begin{eqnarray}
\Delta S &=& \left[\frac{\partial S}{\partial T}\right]_{\begin{subarray}{l}T=T_0,\\
    V_s = V_0\end{subarray}} \Delta T + \left[\frac{\partial S}{\partial V_s}\right]_{\begin{subarray}{l}T=T_0,\\
    V_s = V_0\end{subarray}} \Delta V_s.\label{eq:adiabatic}
\end{eqnarray}
For adiabatic changes, $\Delta S = 0$ and noting that 
$T\left[\frac{\partial S}{\partial T}\right]_{T=T_0,V_s = V_0} = c_v(T=T_0,V_s = V_0)$, where $c_v$ is the volume
heat capacity and depends on the initial temperature and the potential difference, we can write 
\begin{eqnarray}
\left[\frac{\Delta T}{\Delta V_s}\right]_{\begin{subarray}{l}T=T_0,\\
    V_s = V_0\end{subarray}} &=& - \frac{e/k_B}{c_v(T=T_0,V_s = V_0)}\left[\frac{\partial S}
{\partial \psi_s}\right]_{\begin{subarray}{l}T=T_0,\\
    \psi_s = \frac{eV_0}{k_B T_0}\end{subarray}}\label{eq:electrocaloric_heat}
\end{eqnarray}
Here, we have defined $\psi_s = e V_s/k_B T_0$, $e$ is the charge of an electron and $k_B$ is the 
Boltzmann constant so that $e/k_B = 1.16\times 10^4$ K/V. 
It is to be noted that electrostriction\cite{dielectric}
effects leading to changes
in the volume of the liquids are not taken into account here and form the basis of multi-caloric materials exhibiting electrocaloric and
mechanocaloric/elastocaloric effects. This is an interesting direction for future research.
Eq. ~\ref{eq:electrocaloric_heat} provides three insights. First, it is clear that the changes in 
temperature resulting from changes in the potential difference are
inversely proportional to the volume heat capacity of the material. Hence, fluids with low 
heat capacity are preferable candidates for enhanced electrocalorics. 
Second, insight is obtained from the use of 
thermodynamic rules stating\cite{hansenbook} that entropy must increase with an increase the temperature i.e., $c_v>0$. This implies 
that the dimensionless quantity $eV_s/k_BT_0$, ratio of the electrostatic energy of a unit charge to the thermal energy, is the relevant 
variable. In particular, sign of changes in the temperature (i.e., increase or decrease) with an increase in the surface potential depends on the changes in 
entropy with respect to $eV_s/k_BT_0$. Third, the length scale of the region undergoing changes in temperature is determined 
by the volume undergoing entropic changes. 

Larger entropic changes resulting from small changes in the potential difference are required for enhancing the 
electrocaloric effect (cf. Eq. ~\ref{eq:electrocaloric_heat}). As larger entropic changes are expected in 
liquids\cite{neese,liquidECE} than
solids in the presence of an external field, we have focused 
on a theoretical description of the electrocaloric effect in electrolyte solutions.
We use Eq. ~\ref{eq:electrocaloric_heat} and entropic changes computed using field theory\cite{fredbook,kumar_kilbey} to study the
electrocaloric effect in planar double layer systems\cite{verweybook,carniereview,intermolecular_forces}. 
The free energy of the double layer can be constructed\cite{chanmitchellfree, dill, overbeekPaper, carniechanfree, biesheuvel}  with 
different approximations including various effects due to dielectric 
saturation\cite{grahamesaturation,grahameunsymmetrical,orland}, finite 
polarizability of ions\cite{hatlo,andelmanloop,andelmancapacitance}, finite size of 
ions\cite{borukhov,kornyshev,bazant,kornyshevreview}, ion adsorption-desorption 
equilibrium\cite{parsegian} and image charges\cite{carniereview,zgwang1}. This allows systematic 
investigations into roles played by different factors in affecting the electrocaloric effect and pave 
the way for rational design of enhanced electrocaloric fluids. Another motivation in studying such 
a system lies in the need for an improved theory for the electrolyte solutions in strong external fields, 
where crowding and dielectric 
saturation effects are important and a larger electrocaloric effect is observed for 
viscoelastic materials such as polymer\cite{neese} films and liquid-crystalline solutions\cite{liquidECE}. 
Furthermore, 
novel technologies\cite{brogioli,roij} for extracting energy by mixing fresh river water with saline ocean water 
can benefit from 
an improved theory for the electric double layer. These technologies are based on the well-known fact that an 
electric double-layer acts as a capacitor and salt concentration plays a key role in dictating its capacitance. Operating temperature has been shown to play a key role in affecting the energy that can be harvested\cite{roij} using these 
technologies.

We use a microscopic field theory approach to study planar
double layer systems (see the Supporting Information). 
In particular, we consider two parallel plates having surface area $A$ each, 
separated by distance $L$ and immersed in 
an electrolyte solution containing  equal number density ($=\rho_{c,b}$)  
of positive and negative ions along with $\rho_{s,b}$ as the number density of solvent molecules. 
The plates are assumed to have \textit{uniform} surface charge densities (number of charges per unit area), 
$\sigma_1$ and $\sigma_2$ and the corresponding surface 
potentials are $V_{0,1}$ and $V_{0,2}$ (in units of Volts), respectively. Surface potentials and charge densities 
are related to each other by electrostatic boundary conditions and depend on the mechanisms 
by which the plates acquire the surface charge. These relations can be formally derived by 
considering different mechanisms for charging. We take 
molecular volumes of the solvent, positive and negative ions to be $v_s,v_+$ and $v_-$, respectively. 
Noting that theoretical description of polarization 
under an external electric field and strong electric fields are pre-requisites for developing theory 
for the electrocaloric effect, field dependent dielectric and steric effects resulting from finite sizes of ions and solvent
molecules are included in our model. 
In this work, we have built a minimal model that can capture the underlying physics 
based on treating each solvent molecule as an electric 
dipole of length $p_s$ occupying molecular volume $v_s$. Finite polarizability of ions and solvent 
molecules are not considered in this work. However, the current formalism can be extended to take into account the 
effects of polarizability. We have used the theory to study the electocaloric effects in non-overlapping double layers (i.e., single double layer systems) 
so that $V_{0,1} = V_0$ and $V_{0,2}=0$ i.e., conditions of constant surface potentials are considered in this work 
so that the potential difference $V_{s}=V_0$. 
Parameters are chosen to describe water molecules (such as the dipole moment $p_s$). Furthermore, in these 
model calculations, we have considered symmetric ions and solvent molecules so that $v_s = v_+ = v_- = a^3$ and ignored the asymmetry in 
sizes of the molecules. The size parameter $a$ is chosen so that the density of pure water is reproduced i.e., $1/a^3 \equiv 1$ gm/cm$^3$. 

\begin{figure}[htbp]  \centering
\vspace{0.3in}
\subfigure[]{\includegraphics[width=3in]{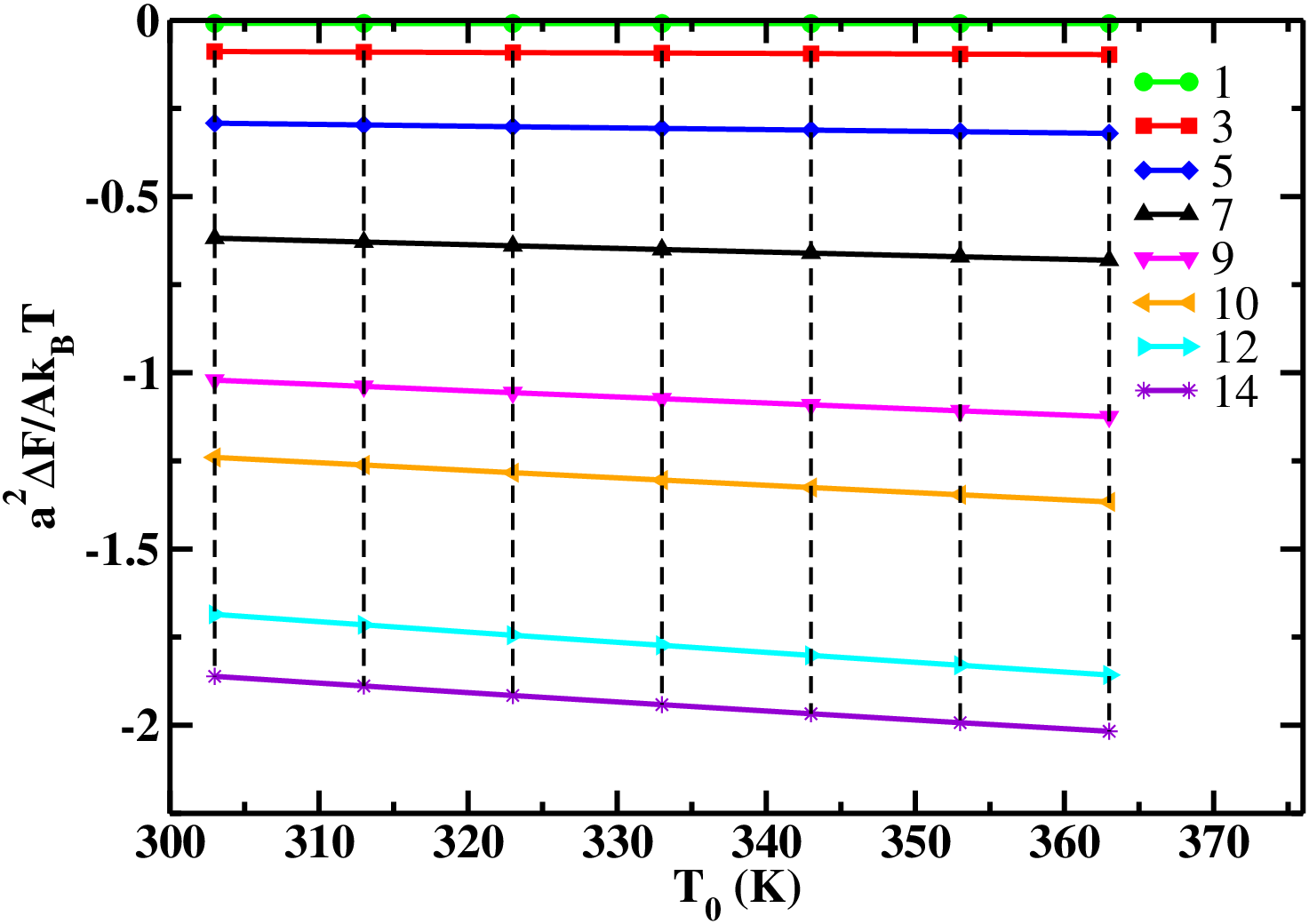}}\vspace{0.26in}
\subfigure[]{\includegraphics[width=3in]{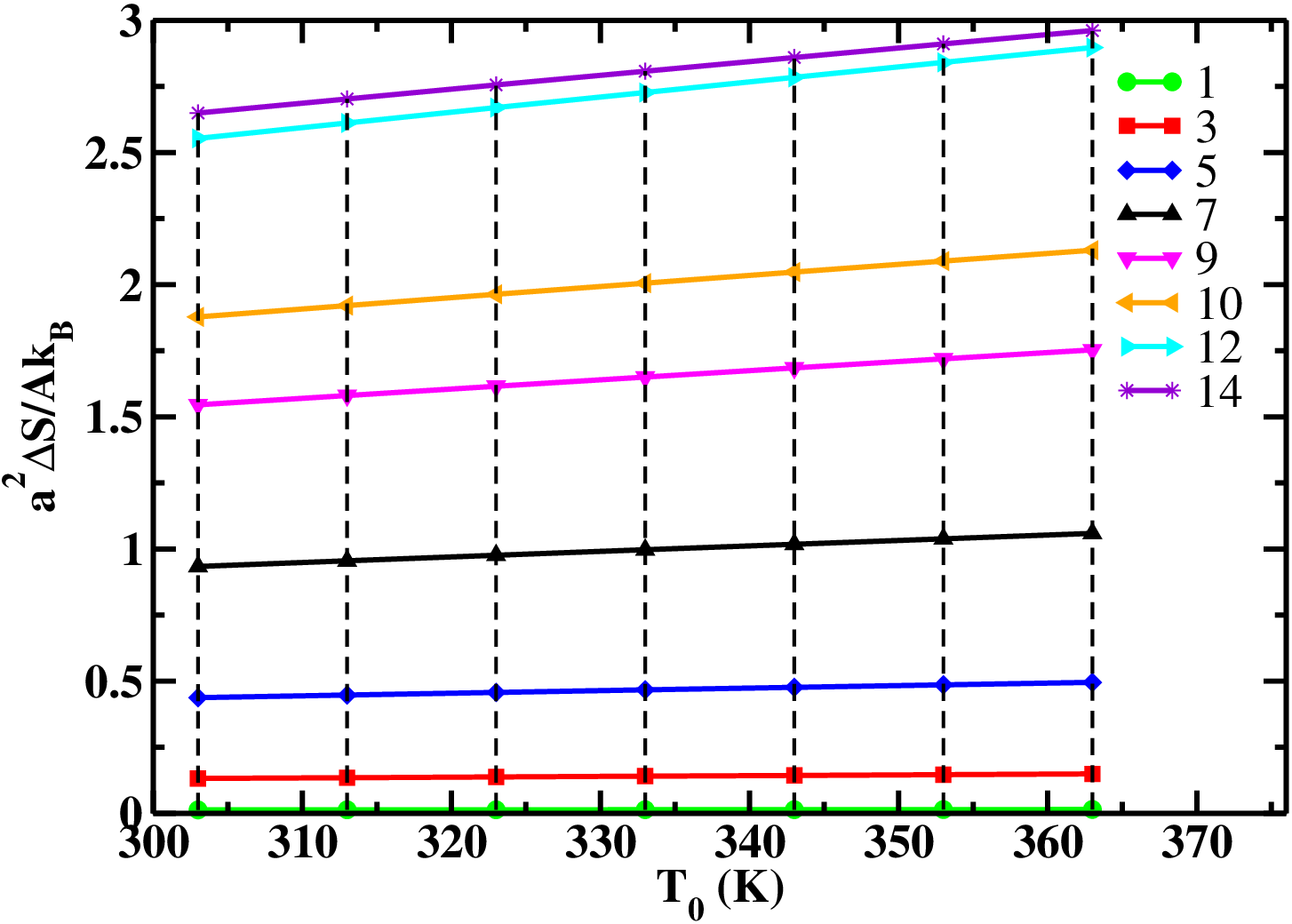}}
\caption{(a) Changes in the free energy and (b) entropy as a function of applied surface potential ($V_0$) and 
temperature for an electolyte solution containing $0.1$ M monovalent salt. Legends show the values of $eV_0/k_B T_0$}
\label{fig:free_energy_entropy_p1}
\end{figure}

Typical free energy changes ($\Delta F$) of the double layers (with respect to the electrolyte solution in the absence of applied 
surface potential ) are shown in Figure ~\ref{fig:free_energy_entropy_p1}(a) for different 
values of $eV_0/k_B T_0$ and temperature ($T_0$) ranging from room temperature to 
near the boiling point of water. The free energy changes are negative for the entire parameter 
range, which is in qualitative agreement with the predictions of the standard Poisson-Boltzmann (PB) approach 
(i.e., ignoring field-dependent dielectric 
and steric effects) and the modified Poisson-Boltzmann (MPB) approach (i.e., ignoring field-dependent dielectric effects) 
(cf. Eqs. $47$ and $43$, respectively, in the Supporting Information). Also, larger free energy 
changes are found with an increase in the temperature due to increased 
entropic contributions shown in Figure ~\ref{fig:free_energy_entropy_p1}(b).
Furthermore, an increase in the free energy changes with an increase in the surface potential 
is also in qualitative agreement with the PB and MPB approaches.
Corresponding entropic changes ($\Delta S = -(\partial \Delta F/\partial T)_{\Omega}, \Omega = AL$ being the total volume) 
such as those shown in Figure ~\ref{fig:free_energy_entropy_p1}(b) 
dictate the electrocaloric effect. 

\begin{figure}[htbp]  \centering
\vspace{0.3in}
\subfigure[]{\includegraphics[width=3in]{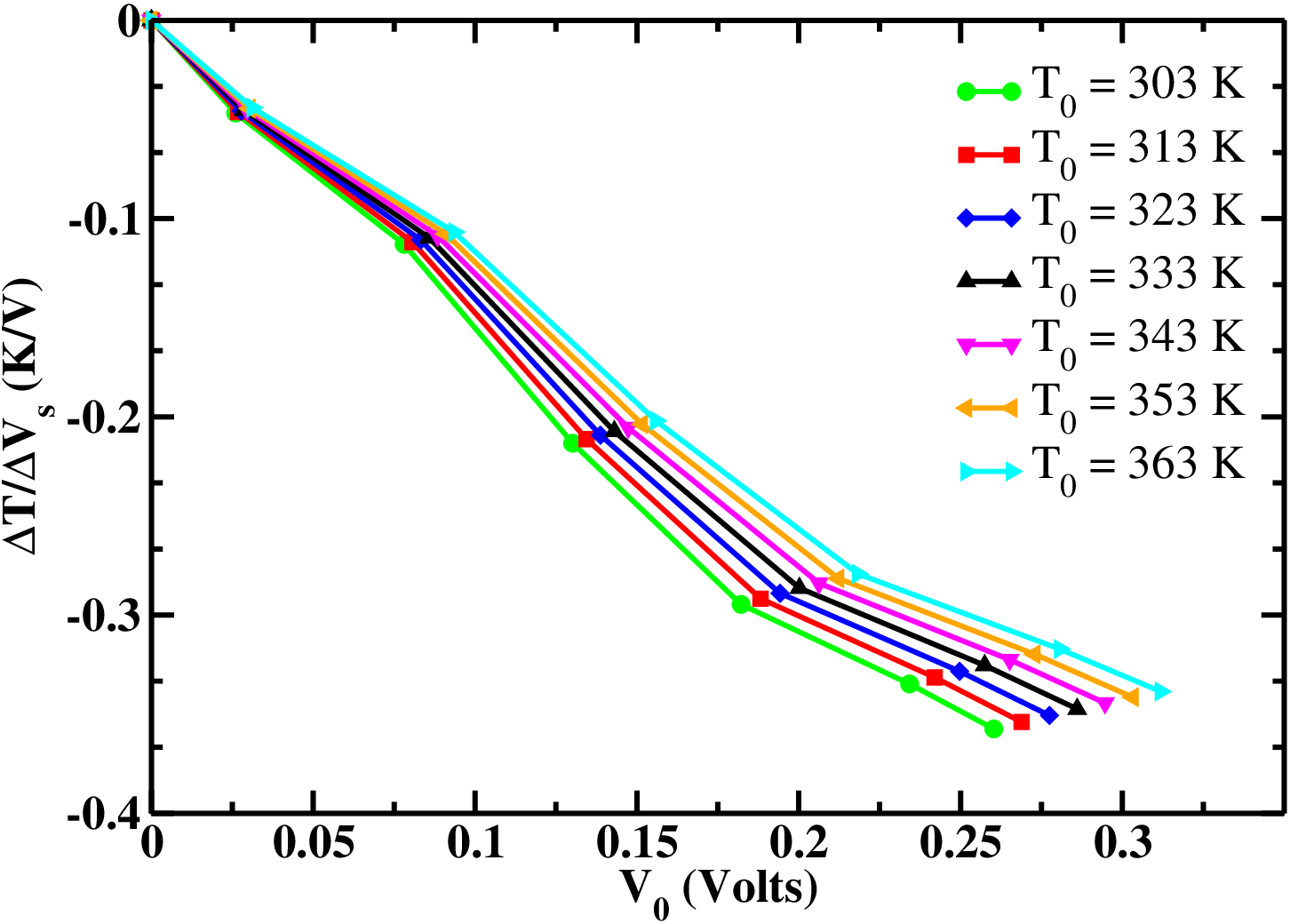}}\vspace{0.26in}
\subfigure[]{\includegraphics[width=3in]{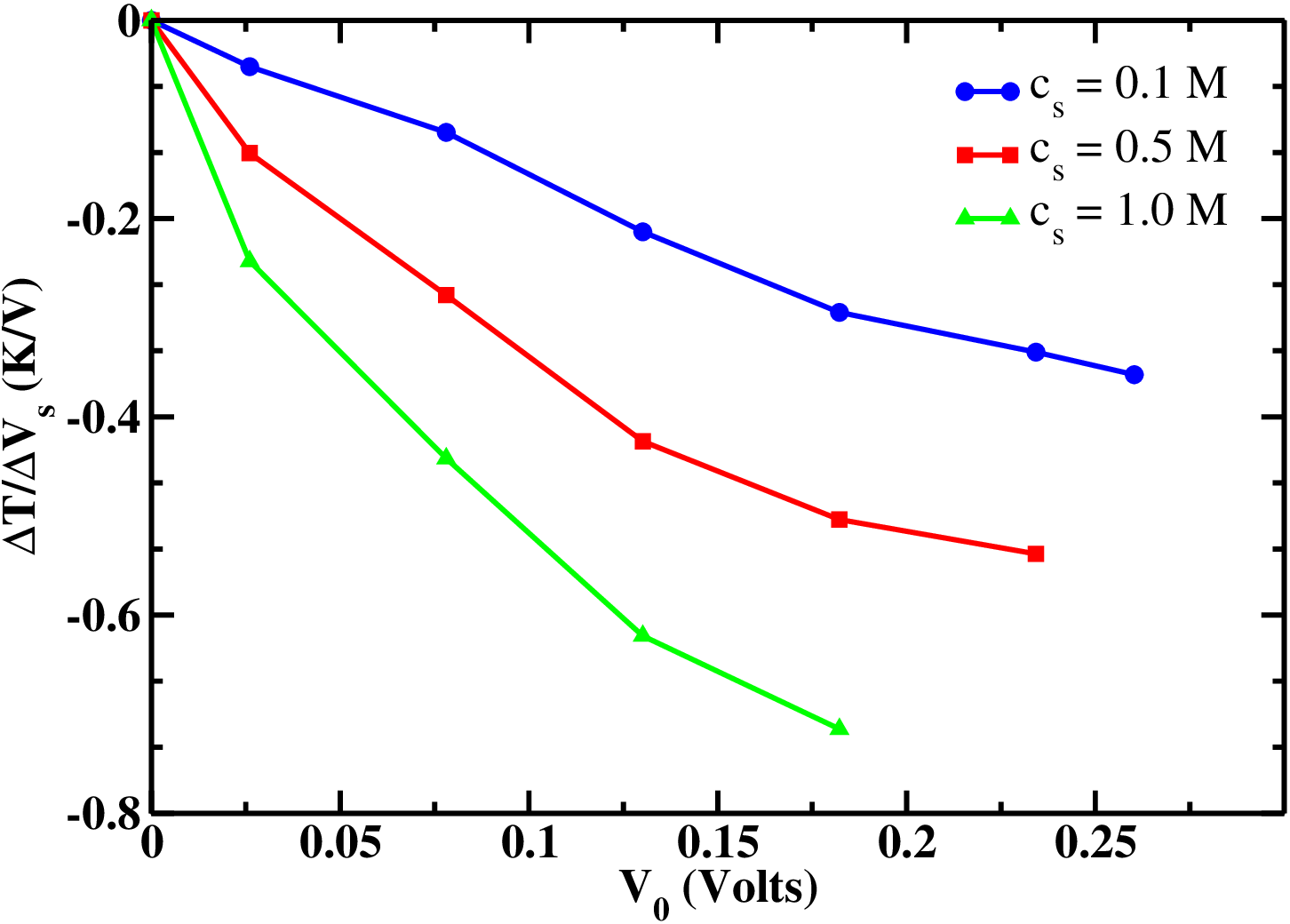}}
\caption{(a) Effects of initial temperature ($T_0$) and (b) the bulk salt concentration (so that $\rho_{c,b} = 0.6023 c_s$ (nm)$^{-3}$ and $c_s$ is in moles per litre (M)) on the electrocaloric effect. The left panel corresponds to $c_s = 0.1$ M and the 
right panel corresponds to $T_0 = 303$ K. Parameter for double layer thickness $L_{dl}/a$ is 
found to be $3,4$ and $6$ for $c_s = 1.0,0.5$ and $0.1$ M, respectively.}
\label{fig:electrocaloric_numerical}
\end{figure}

As the free energy and entropy \textit{ changes per unit area} are computed 
using the field theory, we rewrite Eq. ~\ref{eq:electrocaloric_heat} to calculate the electrocaloric effect so that
\begin{eqnarray}
\left[\frac{\Delta T}{\Delta V_s}\right]_{\begin{subarray}{l}T=T_0,\\
    V_s = V_0\end{subarray}} &=& - \frac{e/k_B}{\bar{c}_v}\left[\frac{\partial \Delta S/\bar{A}k_B}{\partial \psi_s}\right]_{\begin{subarray}{l}T=T_0,\\
    \psi_s = \frac{eV_0}{k_B T_0}\end{subarray}}\label{eq:electrocaloric_heat_2}
\end{eqnarray}
where $\bar{A} = A/a^2$ and 
$\bar{c}_v = c_v/\bar{A}k_B$ is the rescaled
heat capacity of the electrolyte solution in the presence of applied electric field. 
Formally, it can be written as $\bar{c}_v = \bar{c}_v(T=T_0,V_s = 0) + \bar{c}_v(T=T_0,V_s = V_0)$ 
so that $\bar{c}_v(T=T_0,V_s = 0) = T\left[\frac{\partial S_h/\bar{A}k_B}{\partial T}\right]_{\begin{subarray}{l}T=T_0,\\
    \psi_s=0\end{subarray}}$ is the rescaled heat capacity of the reference homogeneous electrolyte solution 
having $S_h$ as its entropy 
and $\bar{c}_v(T=T_0,V_s = V_0) = T\left[\frac{\partial \Delta S/\bar{A}k_B}{\partial T}\right]_{\begin{subarray}{l}T=T_0,\\
    \psi_s=\frac{eV_0}{k_B T_0}\end{subarray}}$ accounts for additional contributions due to the applied electric field. 
It is to be noted that in 
Figure ~\ref{fig:free_energy_entropy_p1}(b), surface potentials and temperature are varied simultaneously due 
to the variation of $\psi_0 = eV_0/k_B T_0$ and the quantity $\frac{\partial \Delta S/\bar{A}k_B}{\partial T}$ can 
be extracted from Figure ~\ref{fig:free_energy_entropy_p1}(b) using the formal relation 
$\frac{\partial \Delta S/\bar{A}k_B}{\partial T} = \left[\frac{\partial \Delta S/\bar{A}k_B}{\partial T}\right]_{\psi_s=\psi_0} - \left[\frac{\psi_0}{T}\frac{\partial 
\Delta S/\bar{A}k_B}{\partial \psi_0}\right]_{T = T_0}$. In calculating the electrocaloric effect, we have taken $\bar{c}_v(T=T_0,V_s = 0) 
= 6.0 L_{dl}/a$ corresponding to molar heat capacities of water to be $3.0 R$ (taken to be independent of temperature) 
and $3/2 R$ for each type of ion treated as an ideal gas\cite{hansenbook} in the homogeneous phase, 
where $L_{dl}$ is the thickness of the double layer and 
$R = k_B N_A$ is the universal gas constant so that $N_A$ is the Avogadro's number. It is to be noted that 
$L_{dl}$ naturally sets the length scale of the region undergoing changes in temperature as a result of the electrocaloric 
effect. For the numerical estimates, we have defined $L_{dl}/a$ as the distance from the electrode after which counterion and coion 
densities approach their bulk values, $\rho_{c,b}$, 

From isothermal changes in the entropy in Figure ~\ref{fig:free_energy_entropy_p1}(b), 
it is clear that entropy of the double layer increases with an increase in the  
surface potential (i.e.,  
$\partial \Delta S/\partial \psi_s > 0$). Such an increase in the entropy 
is in qualitative agreement with the predictions based on PB and MPB approaches (see  
Eqs. $48$ and $46$,respectively, in the Supporting Information). 
As per Eq. ~\ref{eq:electrocaloric_heat_2}, this should lead to 
a decrease of temperature with an increase in the surface potential i.e., 
a negative electrocaloric effect is expected. 
Indeed, such a behavior is observed in Figure ~\ref{fig:electrocaloric_numerical} for 
different initial temperatures and salt concentrations in the bulk. 
Figure ~\ref{fig:electrocaloric_numerical} provides the magnitude of the electrocaloric effect. As an example, consider an electrolyte solution 
containing $1.0$ M monovalent salt with an electrode at surface potential of $0.2$ V at $T_0 = 303$ K 
(near room temperature). For this particular system, $\Delta T/\Delta V_s \sim -0.8 $ K/V is determined from Figure ~\ref{fig:electrocaloric_numerical}(b)
so that a temperature decrease of $0.16$ K is predicted. 
It is found that
magnitude of $\Delta T/\Delta V_s$ is dependent on the initial temperature and the salt concentration in
the bulk. In particular, the magnitude decreases with an increase in
the temperature and increases with an increase in the salt concentration. The decrease in the magnitude with an
increase in the initial temperature is a direct outcome of an increase in the heat capacity of the double layer
with an increase in the applied surface potential, as evident from Figure ~\ref{fig:free_energy_entropy_p1}(b).
The increase in the magnitude of the electrocaloric effect with an increase in the 
bulk salt concentration results from a decrease in thickness of the double layer ($L_{dl}$). Furthermore, 
larger free energy and entropic changes are found with an increase in the bulk salt concentration, 
as shown in Figure $1$(a) in the Supporting Information. 
It should be noted that qualitatively the same effects are predicted by the PB and MPB approaches, where the free energy and 
entropy changes increase as $\sqrt{\rho_{c,b}}$. However, quantitatively, the PB and MPB approaches digress from the 
numerical results due to errors made in predicting the free energy changes. To demonstrate this point, we have 
shown a comparison of the free energy changes for the same system, estimated using the PB, MPB and the numerical calculations in 
Figure ~\ref{fig:free_energy_all}(a). It is found that the PB approach 
is off by factors of $10-100$ for $eV_0/k_B T_0 > 10$. In contrast, the MPB approach 
corrects for some of the errors made in the PB approach but it still deviates from the numerical results by a factor of $3$.  

\begin{figure}[htbp]  \centering
\vspace{0.3in}
\subfigure[]{\includegraphics[width=3in]{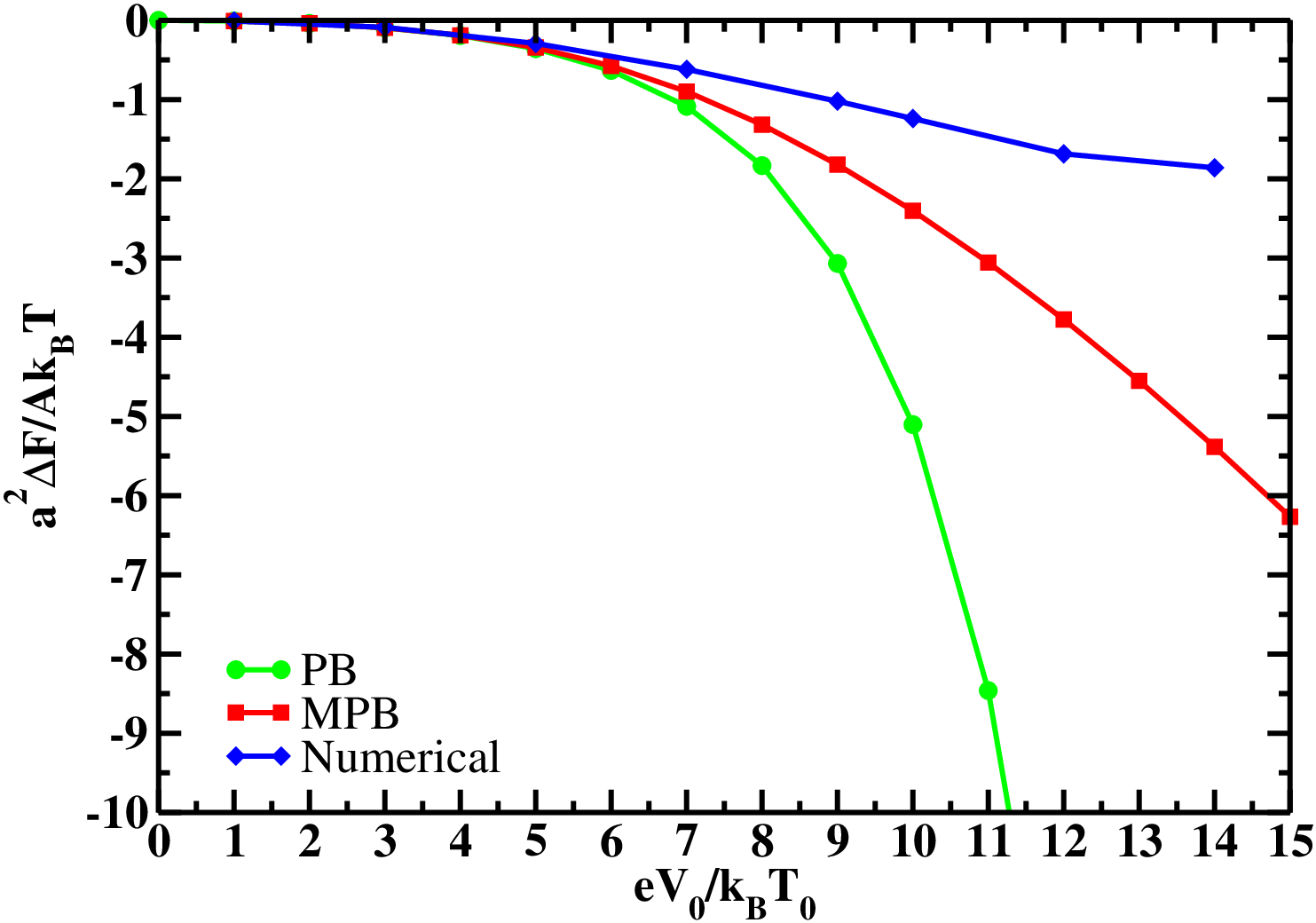}}\vspace{0.26in}
\subfigure[]{\includegraphics[width=3in]{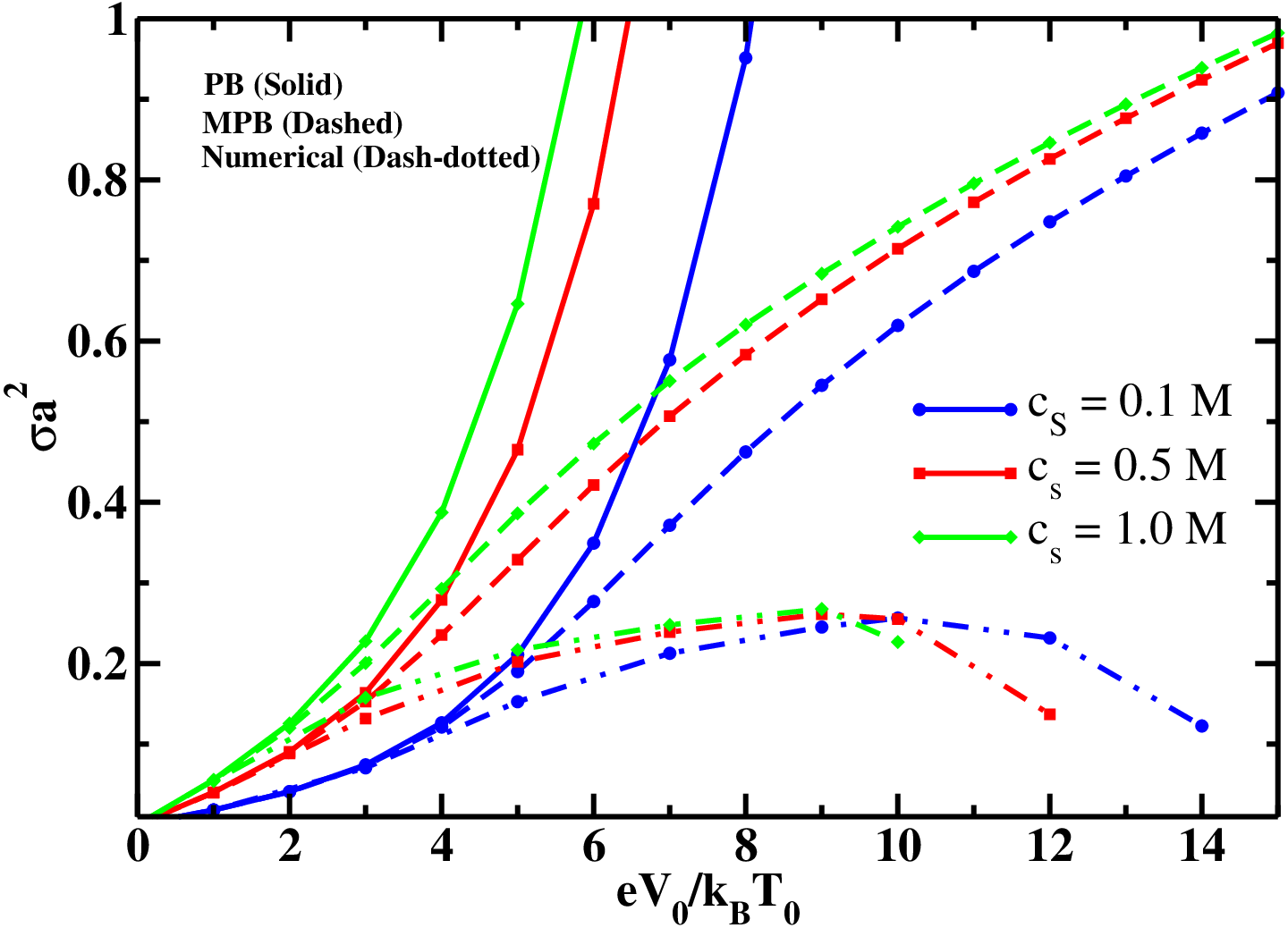}}
\caption{(a) Comparison of the free energy changes computed using the PB approach (i.e., ignoring field-dependent dielectric and steric effects), 
the MPB approach (i.e., ignoring field-dependent dielectric effects) and the numerical calculations for $c_s = 0.1$ M, $T_0 = 303$ K. 
(b) Computed surface charge density as a function of applied surface potential, estimated using the PB, the MPB
and the numerical calculations, for different bulk salt concentrations at $T_0 = 303$ K. The solid lines correponds to
the analytical relation presented in the text.}
\label{fig:free_energy_all}
\end{figure}
 
The relative importance of the field-dependent dielectric and steric effects in
predicting structure of the double layer and resulting changes in the free energy can be assessed by 
comparing plots showing surface charge density as a function of the surface potential as predicted by 
the PB, MPB and numerical approaches (Figure ~\ref{fig:free_energy_all}(b)). The surface charge density in the PB approach is given by
$\sigma_1 = \sigma = \frac{\kappa \epsilon_h}{2\pi |Z_c|l_{Bo}}\sinh\left[\frac{|Z_c| eV_0}{2k_B T}\right]$
where $|Z_c|$ is the valency of ions ($=1$ for monovalent ions), $l_{Bo} = e^2/4\pi\epsilon_0 k_B T$ so that $\epsilon_0$ is the
permittivity of vacuum, $\epsilon_h = 1 + 4\pi l_{Bo} p_s^2 \rho_{s,b}/3$ is the relative permittivity of the homogeneous
electrolyte solution so that $\rho_{s,b}$ is the solvent density
and $\kappa = (8\pi l_{Bo} |Z_c|^2 \rho_{c,b}/\epsilon_h)^{1/2}$ is the inverse Debye screening length.
The PB and MPB approaches predict a monotonic increase in the surface charge density with an increase in the surface potential, as shown in 
Figure ~\ref{fig:free_energy_all}(b),
without showing any sign of saturation, leading to unphysical surface charge densities. 
The numerical calculations
show agreement with the PB and MPB approaches for $eV_0/k_B T_0 < 2-4$ depending on the salt concentration and deviate strongly
for higher surface potentials exhibiting saturation and a decrease in the surface charge density.
This is an outcome of dielectric saturation leading to lowering of dielectric function near the surface, 
ignored in the PB and MPB approaches. 
The decrease in the surface charge
density with an increase in the surface potential (i.e., negative differential capacitance\cite{negdiff1,negdiff2}, where
 differential capacitance $=\partial \sigma/\partial V_0$) hints at
the breakdown of the
one-dimensional uniform charge density model used here and plausible onset of in-plane charge density waves\cite{negdiff1}. 


In conclusion, we have presented a field theory approach for studying electrocaloric effects in planar double layer systems. 
Two key ingredients of the theory are the consideration of steric effects and
dipolar interactions resulting from polar solvent molecules. 
Although the theory is general, in 
this work, we have presented calculations 
for aqueous solutions containing monovalent salt ions. It was shown 
that the electrocaloric 
effect in planar double layer systems is negative, i.e., the temperature of the double layer should decrease with an 
increase in the applied surface potential. The magnitude of the electrocaloric effect depends on the 
initial temperature of the solution and the salt concentration. In particular, we showed 
that the magnitude of the electrocaloric effect should decrease with increase in the initial temperature and 
increase with an increase in the salt concentration. 
Due to the general nature of the field theory approach\cite{fredbook} to tackle curved interfaces, 
polymers, multivalent ions etc., our work opens up a new area of theoretical research focused on 
the rational design of electrocaloric fluids. Furthermore, we have shown that the field theory approach 
stays robust for high surface potentials and the other approaches such as the 
PB and MPB are not reliable. This particular feature of the field theory is quite important 
for energy harvesting technologies based on electrochemical capacitors and supercapacitors. 

We acknowledge support from the Center for Nanophase 
Materials Sciences, which is sponsored at Oak Ridge National Laboratory by the Scientific User Facilities Division, office of Basic Energy Sciences, U.S. Department of Energy (DOE). 

\vspace{-0.2in}

\section{Supporting Information: Theory}\label{sec:theory}
\setcounter {equation} {0}
We consider two parallel plates separated by distance $L$ and immersed in
an electrolyte solution containing $n_s$ solvent molecules, $n_{+}$ positive and $n_{-}$ negative ions.
The plates are assumed to have \textit{uniform} surface charge densities (number of charges per unit area),
$\sigma_1$ and $\sigma_2$ (in units of electronic charge, $e$) and the corresponding surface
potentials are $V_{0,1}$ and $V_{0,2}$, respectively. It is to be noted that surface potentials and charge densities
are related to each other by electrostatic boundary conditions\cite{intermolecular_forces} and depend on the mechanisms
by which the plates acquire the surface charge. These relations can be formally derived by
considering different mechanisms for charging\cite{intermolecular_forces}.

Molecular volumes of the solvent, positive and negative ions are taken to be $v_s,v_+$ and $v_-$, respectively.
We are interested in understanding the effects of dipolar interactions and finite ion sizes on the thermodynamics of
double layer. For such purposes, we seek a minimal model that can capture the underlying physics.
In this work, we have studied a minimal model based on treating each solvent molecule as an electric
dipole of length $p_s$ occupying molecular volume $v_s$. Also, the positive and negative ions
have molecular volumes of $v_+$ and $v_-$, respectively. Finite polarizability\cite{onsager_moments,dielectric}
of ions and solvent
molecules are not considered in this work. However, the current formalism can be extended to take into account the
effects of polarizability.

The canonical partition function for such a system is written\cite{orland,kumar_kilbey} as
\begin{eqnarray}
       Z & = & \int \prod_{j=\pm,s}\frac {1}{n_j!}\prod_{\alpha=1}^{n_j} d\mathbf{r}_{j,\alpha} \int \prod_{\alpha=1}^{n_s} d\mathbf{u}_{\alpha} \exp \left [
- \hat{H}\left\{\mathbf{r}_{j,\alpha},\mathbf{u}_{\alpha}\right\} 
\right ] \prod_{\mathbf{r}}\mathbf{\delta}\left(\sum_{j=\pm,s}\hat{\rho}_{j}(\mathbf{r})v_j - 1\right)\label{eq:parti_solutions}
\end{eqnarray}
where $\mathbf{r}_{j,\alpha}$ is the position vector for the $\alpha^{th}$ particle of type $j$
and $\mathbf{u}_{\alpha}$ is the unit vector quantifying orientation of $\alpha^{th}$ solvent dipole.
The Hamiltonian is written by taking into account
the contributions coming from ion-ion, ion-dipole and dipole-dipole interactions.
Short range interactions between ions and solvent molecules are ignored in the minimal model studied here.
$\hat{\rho}_{j}(\mathbf{r})$ represents microscopic number
density of the particles of type $j$ at a certain location $\mathbf{r}$ defined as
        \begin{eqnarray}
\hat{\rho}_{j}(\mathbf{r})  &=& \sum_{\alpha=1}^{n_j} \delta \left(\mathbf{r}-\mathbf{r}_{j,\alpha}\right) 
\quad \mbox{for} \quad j= s,+,-
\end{eqnarray}
The Hamiltonian for the ions and dipoles can be written as\cite{orland,kumar_kilbey}
\begin{eqnarray}
\hat{H} &=& \frac{l_{Bo}}{2}\int d\mathbf{r}\int d\mathbf{r}' 
\frac{\left[\hat{\rho}_e(\mathbf{r})-\nabla_{\mathbf{r}}.\hat{P}(\mathbf{r}) \right]\left[\hat{\rho}_e(\mathbf{r}')-\nabla_{\mathbf{r}'}.\hat{P}(\mathbf{r}')
\right]}{|\mathbf{r} - \mathbf{r}'|} \label{eq:final_particle_elec}
\end{eqnarray}
where $l_{Bo} = e^2/4\pi\epsilon_o k_B T$ is the Bjerrum length in vaccum and $\hat{\rho}_e(\mathbf{r})$ is the
charge density (in units of $e$), given by
$\hat{\rho}_e(\mathbf{r}) = \sum_{j=\pm}Z_j \hat{\rho}_{j}(\mathbf{r}) + \sigma_1 \delta(x-x_1) 
+ \sigma_2 \delta(x-x_2)$, $Z_{j}$ being the
valency (with sign) of ions of type $j$ and $|x_2-x_1| = L$ is the distance between the plates.
Also, $\hat{P}(\mathbf{r})$ is polarization density of dipoles (in units of $e$) at location
$\mathbf{r}$, given by
\begin{eqnarray}
\hat{P}(\mathbf{r})  &=& p_s \sum_{\alpha=1}^{n_s} \delta \left(\mathbf{r}-\mathbf{r}_{\alpha}\right) \mathbf{u}_{\alpha}
\end{eqnarray}

\subsection{Field theory in the canonical ensemble}
A field theory for the system described above can be constructed following a standard
protocol\cite{fredbook}. We start from the electrostatics contributions to the partition function.
For the electrostatics contribution to the partition function written in the form
Eq. ~\ref{eq:final_particle_elec}, we use Hubbard-Stratonovich transformation\cite{fredbook} so that
\begin{eqnarray}
\exp\left[-\hat{H}\right] &=& \frac{1}{N_\psi} \int D\left[\psi\right]\exp \left[-i\int d\mathbf{r}
\left\{\hat{\rho}_{e}(\mathbf{r}) - \nabla_{\mathbf{r}}.\hat{P}(\mathbf{r})\right\} \psi(\mathbf{r}) +
\frac{1}{8\pi l_{Bo}}\int d\mathbf{r}\psi(\mathbf{r})\nabla_\mathbf{r}^2 \psi(\mathbf{r}))\right] \nonumber \\
&&
\end{eqnarray}
where $N_\psi$ is a normalization factor, given by
\begin{eqnarray}
N_\psi &=& \int D\left[\psi\right]\exp \left[\frac{1}{8\pi
l_{Bo}}\int d\mathbf{r}\psi(\mathbf{r})\nabla_\mathbf{r}^2 \psi(\mathbf{r}))\right]
\end{eqnarray}
Using this transformation and writing
the local constraints (represented by delta functions) in terms of functional integrals using
\begin{eqnarray}
\prod_{\mathbf{r}}\mathbf{\delta}\left(\sum_{j=\pm,s}\hat{\rho}_j(\mathbf{r})v_j - 1\right)
 &=& \int D\left[\eta\right] \exp\left[-i \int d\mathbf{r} \eta(\mathbf{r})\left\{\sum_{j=\pm,s}\hat{\rho}_j(\mathbf{r})v_j - 1\right\}\right] \label{eq:order_field}
\end{eqnarray}
we can write the partition function given by Eq. ~\ref{eq:parti_solutions} as
\begin{eqnarray}
       Z & = & \frac{1}{N_\psi}\int D\left[\psi\right]\int  D\left[\eta\right]
\exp \left [- \frac{H}{k_BT}\right ] \label{eq:parti_physical}
\end{eqnarray}
where
\begin{eqnarray}
       \frac{H}{k_BT} &=& - \frac{1}{8\pi l_{Bo}}\int d\mathbf{r} \psi(\mathbf{r})\nabla_{\mathbf{r}}^2 \psi(\mathbf{r})
- i \int d\mathbf{r} \eta (\mathbf{r}) + \sigma_1 \int d\mathbf{r_{\parallel}}i \psi(\mathbf{r_{\parallel}},x_1) \nonumber \\
&& + \sigma_2 \int d\mathbf{r_{\parallel}}i \psi(\mathbf{r_{\parallel}},x_2)
- \sum_{j=\pm,s}\left\{n_j \ln Q_{j}\left\{\psi,\eta\right\} - \ln n_j!\right\}\label{eq:hami_physical}
\end{eqnarray}
and we have used the notation $\mathbf{r} = (x,y,z)\equiv (x,\mathbf{r_{\parallel}})$ so that $\mathbf{r_{\parallel}}$
denotes in-plane vector parallel to the plates. $Q_{j}$ is the partition function for particles of type $j$, given
by
\begin{eqnarray}
      Q_{j=\pm}\left\{\psi,\eta\right\} &=& \int d \mathbf{r} 
\exp\left[-iZ_j\psi(\mathbf{r})-iv_j \eta(\mathbf{r})\right] \\
Q_{s}\left\{\psi,\eta\right\} &=& \int d \mathbf{r} \int d\mathbf{u}  
\exp\left[-ip_s \mathbf{u}.\nabla_{\mathbf{r}}\psi(\mathbf{r})-iv_s \eta(\mathbf{r})\right] 
\end{eqnarray}

In the following, we
use the saddle-point approximation to estimate the functional integrals over $\psi$ and $\eta$.
An equivalent calculation in the grand canonical ensemble is presented in the Appendix A.

\subsection{Saddle-point approximation: self-consistent equations, free energy and chemical potentials}
The saddle point approximation with respect to $\eta$ and $\psi$ gives two non-linear equations. At the saddle-points,
both $\eta$ and $\psi$ turn out to be purely imaginary. Writing $i\eta(\mathbf{r}) = \eta^\star(\mathbf{r})$
and $i\psi(\mathbf{r}) = \psi^\star(\mathbf{r})$ at the saddle point and
\textit{defining} densities of ions and solvent molecules via
\begin{eqnarray}
       \rho_{j=\pm}(\mathbf{r}) &=& \frac{n_j}{Q_{j}\left\{\psi^{\star},\eta^{\star}\right\}} \exp \left [- Z_j \psi^{\star}(\mathbf{r}) - v_j\eta^\star(\mathbf{r})\right ] \label{eq:den_ions}\\
   \rho_{s}(\mathbf{r}) &=&
        \frac{4\pi n_s}{Q_{s}\left\{\psi^{\star},\eta^{\star}\right\}} \exp\left[- v_s \eta^\star(\mathbf{r})\right] \frac{\sinh p_s |\nabla_{\mathbf{r}}\psi^{\star}(\mathbf{r})|}{p_s |\nabla_{\mathbf{r}}\psi^{\star}(\mathbf{r})|} \label{eq:den_solvent}
\end{eqnarray}
the equations at the saddle point are given by
\begin{eqnarray}
       \sum_{j=\pm} v_j \rho_j(\mathbf{r}) + v_s \rho_s(\mathbf{r}) &=& 1 \label{eq:saddle1}\\
\nabla_{\mathbf{r}}\cdot\left[\epsilon(\mathbf{r})\nabla_{\mathbf{r}}\psi^{\star}(\mathbf{r})\right] &=& 
-4\pi l_{Bo}\rho_e(\mathbf{r})\label{eq:saddle2}
\end{eqnarray}
so that the local charge density ($\rho_e(\mathbf{r})$) and dielectric function ($\epsilon(\mathbf{r})$) are given by
\begin{eqnarray}
\rho_e(\mathbf{r}) &=& \sum_{j=\pm}Z_j \rho_j(\mathbf{r}) + \sigma_1 \delta(x-x_1) + \sigma_2 \delta(x-x_2)\\
       \epsilon(\mathbf{r}) &=& 1 + 4\pi l_{Bo}p_s^2 \rho_{s}(\mathbf{r}) \frac{L\left[p_s |\nabla_{\mathbf{r}}\psi^{\star}(\mathbf{r})|\right]}{p_s |\nabla_{\mathbf{r}}\psi^{\star}(\mathbf{r})|}  
\end{eqnarray}
where $L(x) = \coth(x) - 1/x$ is the Langevin function. Corresponding Helmholtz free
energy ($F$) is given by the approximation $F/k_B T = -\ln Z \simeq H^{\star}/k_B T = F^{\star}/k_B T$
so that (cf. Eq. ~\ref{eq:hami_physical})
\begin{eqnarray}
       \frac{F^{\star}}{k_BT} &=& \frac{1}{8\pi l_{Bo}}\int d\mathbf{r} \psi^{\star}(\mathbf{r})\nabla_{\mathbf{r}}^2 \psi^{\star}(\mathbf{r})
- \int d\mathbf{r} \eta^{\star} (\mathbf{r}) + \sigma_1 \int d\mathbf{r_{\parallel}}\psi^{\star}(\mathbf{r_{\parallel}},x_1) \nonumber \\
&& + \sigma_2 \int d\mathbf{r_{\parallel}}\psi^{\star}(\mathbf{r_{\parallel}},x_2)
- \sum_{j=\pm,s}\left\{n_j \ln Q_{j}\left\{\psi^{\star},\eta^{\star}\right\} - \ln n_j!\right\} \label{eq:free_saddle}
\end{eqnarray}
Eq. ~\ref{eq:free_saddle} can be rewritten after eliminating $n_j$ using
Eqs. ~\ref{eq:saddle1} and ~\ref{eq:saddle2}. Furthermore, using the Stirling
approximation $\ln n! \simeq n\ln n -n$, Eq. ~\ref{eq:free_saddle} can be written as
\begin{eqnarray}
       \frac{F^{\star}}{k_BT} &=& \int d\mathbf{r} \rho_e(\mathbf{r})\psi^{\star}(\mathbf{r}) + \frac{1}{8\pi l_{Bo}}\int d\mathbf{r} \psi^{\star}(\mathbf{r})\nabla_{\mathbf{r}}^2 \psi^{\star}(\mathbf{r})\nonumber \\
&& -\int d\mathbf{r} \rho_s(\mathbf{r})\ln \left[4\pi\frac{\sinh p_s |\nabla_{\mathbf{r}}\psi^{\star}(\mathbf{r})|}{p_s |\nabla_{\mathbf{r}}\psi^{\star}(\mathbf{r})|}\right]
+ \sum_{j=\pm,s}\int d\mathbf{r}\rho_j(\mathbf{r}) \left[\ln \rho_{j}(\mathbf{r})-1\right] \label{eq:free_saddle_den}
\end{eqnarray}

For study of opposing double layer systems in equilibrium with an electrolyte solution,
chemical potential is determined by conditions in the solution far from the plates. In order to fix the
chemical potentials by specifying
different conditions in the solution far from the plates, we rewrite the above equations in terms of
chemical potenials. An approximation for the chemical potentials ($\mu_j$) of different species can be derived from
Eq. ~\ref{eq:free_saddle} using the thermodynamic relation $\mu_j = (\partial F/\partial n_j)_{\Omega}
\simeq (\partial F^{\star}/\partial n_j)_{\Omega} = \mu_j^{\star}, \Omega$ being the total volume. Using the Stirling
approximation $\ln n! \simeq n\ln n -n$, the chemical potentials within the saddle-point
approximation are given by
\begin{eqnarray}
       \frac{\mu_{j = \pm, s}^{\star}}{k_B T}  &=& \ln\left[\frac{n_j}{Q_{j}\left\{\psi^{\star},\eta^{\star}\right\}}\right] \label{eq:chem_potential}
\end{eqnarray}
Using Eq. ~\ref{eq:chem_potential}, Eqs. ~\ref{eq:den_ions} and ~\ref{eq:den_solvent} can be written as
\begin{eqnarray}
       \rho_{j=\pm}(\mathbf{r}) &=& \exp \left [\frac{\mu_j^{\star}}{k_B T}- Z_j \psi^{\star}(\mathbf{r}) - v_j\eta^\star(\mathbf{r})\right ] \label{eq:den_ions_g}\\
   \rho_{s}(\mathbf{r}) &=&
        4\pi \exp\left[\frac{\mu_s^{\star}}{k_B T}- v_s \eta^\star(\mathbf{r})\right] \frac{\sinh p_s |\nabla_{\mathbf{r}}\psi^{\star}(\mathbf{r})|}{p_s |\nabla_{\mathbf{r}}\psi^{\star}(\mathbf{r})|} \label{eq:den_solvent_g}
\end{eqnarray}

\subsection{Chemical part of the free energy: charging the electrodes and adsorption-desorption electrochemical equilibrium}
The free energy (cf. Eq. ~\ref{eq:free_saddle_den}) for the two opposing double layer system is
obtained for a \textit{given} surface charge density of the plates and has the charged plates at given surface potentials (in vacuum) as the reference frame. This can be easily seen by putting $\rho_{j=\pm,s} = 0$ in Eq. ~\ref{eq:free_saddle_den} so that
$F^{\star}/k_B T \left\{\rho_{j=\pm,s} = 0\right\} = (\sigma_1/2) \int d\mathbf{r_{\parallel}}\psi^{\star}(\mathbf{r_{\parallel}},x_1) + (\sigma_2/2) \int d\mathbf{r_{\parallel}}\psi^{\star}(\mathbf{r_{\parallel}},x_2)$.
This, in turn, means that Eq. ~\ref{eq:free_saddle_den} doesn't include the work done (typically by an external source) in charging the two plates at a separation distance of $L = |x_1-x_2|$. This contribution\cite{verweybook,overbeekPaper} to the free energy is
\begin{eqnarray}
\frac{F_{chem}}{k_B T} &=& -\int d\mathbf{r_{\parallel}} \int_{0}^{\sigma_1} d\sigma' \psi^{\star}(\mathbf{r_{\parallel}},x_1)\left\{\sigma'\right\} -\int d\mathbf{r_{\parallel}} \int_{0}^{\sigma_2} d\sigma' \psi^{\star}(\mathbf{r_{\parallel}},x_2)\left\{\sigma'\right\} \label{eq:chemical_free}
\end{eqnarray}
Evaluation of the right hand side in Eq. ~\ref{eq:chemical_free} requires specification of the mechanisms by
which the plates acquire their charge. In the following, we consider the specific case when plates are kept at
constant surface potentials.

\subsection{One dimensional model: plates at constant surface potentials with symmetrical ions and solvent molecules}
If the densities far from the plates are known to be $\rho_{j,b}$ corresponding to spatially uniform
$\psi^{\star}(\mathbf{r}) = \psi^{\star}_b$ and
$\eta^{\star}(\mathbf{r}) = \eta^{\star}_b$ then Eqs. ~\ref{eq:den_ions_g} and ~\ref{eq:den_solvent_g} can be written as
\begin{eqnarray}
       \rho_{j=\pm}(\mathbf{r}) &=& \rho_{j,b}\exp \left [- Z_j \left\{\psi^{\star}(\mathbf{r}) -\psi^{\star}_b\right\}- v_j\left\{\eta^\star(\mathbf{r})-\eta^{\star}_b\right\}\right ] \label{eq:den_ions_results}\\
   \rho_{s}(\mathbf{r}) &=&
        \rho_{s,b} \exp\left[- v_s \left\{\eta^\star(\mathbf{r})-\eta^{\star}_b\right\}\right] \frac{\sinh p_s |\nabla_{\mathbf{r}}\psi^{\star}(\mathbf{r})|}{p_s |\nabla_{\mathbf{r}}\psi^{\star}(\mathbf{r})|} \label{eq:den_solvent_results}
\end{eqnarray}

For two parallel plates, saddle point value of $\psi$ varies only along the direction perpendicular to the
charged surface (taken to be along x-axis) so that $\psi^\star(\mathbf{r}) \equiv \psi^\star(x), 
\eta^\star(\mathbf{r}) \equiv \eta^\star(x)$. Furthermore, considering the case of symmetric ions and solvent molecules
so that $v_{j=\pm,s} = a^3$ and $Z_{+} = -Z_- = |Z_c|$ so that $\rho_{j=\pm,b} = \rho_{c,b}$,
we can eliminate $\eta^{\star}$ using Eqs. ~\ref{eq:saddle1},
~\ref{eq:den_ions_results} and ~\ref{eq:den_solvent_results} and write Eq. ~\ref{eq:saddle2} as
\begin{eqnarray}
\frac{\partial }{\partial x}\left[\epsilon(x)\frac{\partial \psi^{\star}(x)}{\partial x}\right] &=& 
-4\pi l_{Bo}\rho_e(x)\label{eq:saddle2_1d}
\end{eqnarray}
where the local charge density ($\rho_e(x)$) and dielectric function ($\epsilon(x)$) are given by
\begin{eqnarray}
\rho_e(x) &=& |Z_c|\left[\rho_+(x)-\rho_-(x)\right] + \sigma_1 \delta(x-x_1) + \sigma_2 \delta(x-x_2)\\
\rho_+(x) &=& \frac{\rho_{c,b}\exp\left[-|Z_c|\left\{\psi^{\star}(x)-\psi^{\star}_b\right\}\right]}{f\left\{\psi^{\star}(x)-\psi^{\star}_b,\frac{\partial \psi^{\star}(x)}{\partial x}\right\}}\\
\rho_-(x) &=& \frac{\rho_{c,b}\exp\left[|Z_c|\left\{\psi^{\star}(x)-\psi^{\star}_b\right\}\right]}{f\left\{\psi^{\star}(x)-\psi^{\star}_b,\frac{\partial \psi^{\star}(x)}{\partial x}\right\}}\\
\epsilon(x) &=& 1 + 4\pi l_{Bo}p_s^2 \frac{\rho_{s,b}}{f\left\{\psi^{\star}(x)-\psi^{\star}_b,\frac{\partial \psi^{\star}(x)}{\partial x}\right\}}\frac{\sinh p_s |\frac{\partial \psi^{\star}(x)}{\partial x}|}{p_s |\frac{\partial \psi^{\star}(x)}{\partial x}|}\frac{L\left[p_s |\frac{\partial \psi^{\star}(x)}{\partial x}|\right]}{p_s |\frac{\partial \psi^{\star}(x)}{\partial x}|} \label{eq:dielec_function}
\end{eqnarray}
so that
\begin{eqnarray}
f\left\{\psi^{\star}(x)-\psi^{\star}_b,\frac{\partial \psi^{\star}(x)}{\partial x}\right\} &=& \rho_{s,b}a^3\frac{\sinh p_s |\frac{\partial \psi^{\star}(x)}{\partial x}|}{p_s |\frac{\partial \psi^{\star}(x)}{\partial x}|} 
+ 2\rho_{c,b}a^3\cosh\left[|Z_c|\left\{\psi^{\star}(x)-\psi^{\star}_b\right\}\right]
\end{eqnarray}
and $\left[\rho_{s,b} + 2\rho_{c,b}\right]a^3 = 1$. It is to be noted that
solvent density is given by
\begin{eqnarray}
\rho_s(x) &=& \frac{\rho_{s,b}}{f\left\{\psi^{\star}(x)-\psi^{\star}_b,\frac{\partial \psi^{\star}(x)}{\partial x}\right\}}\frac{\sinh p_s |\frac{\partial \psi^{\star}(x)}{\partial x}|}{p_s |\frac{\partial \psi^{\star}(x)}{\partial x}|}\label{eq:solv_den}
\end{eqnarray}
and satisfies the incompressibility constraint $\sum_{j=\pm,s}\rho_j(x)a^3 = 1$.

\subsection{Free energy within saddle-point approximation : adiabatic changes}
Changes in entropy ($\Delta S$) can be readily calculated
from the corresponding free energy changes ($\Delta F$) and the thermodynamic relation
$\Delta S = -\left(\frac{\partial \Delta F}{\partial T}\right)_{\Omega}$. Free energy of the double layer system ($F^{\star}_{dl}$) is
the sum of electrostatic contributions approximated by $F^{\star}$ and the
chemical part given by $F_{chem}^{\star}$. Superscript $\star$ implies the use of saddle-point
approximation (mean-field like treatment) in estimating the free energy.
In particular, assuming lateral homogeneity, for plates (at known
surface potentials) separated by distance $L$ having surface area $A$ each, $F^{\star}$ and $F_{chem}^{\star}$
are given by
\begin{eqnarray}
       \frac{F^{\star}}{A k_BT} &=& \int_0^{L} dx \rho_e(x)\psi^{\star}(x) + \frac{1}{8\pi l_{Bo}}\int_0^L dx \psi^{\star}(x)\frac{\partial^2\psi^{\star}(x)}{\partial x^2}\nonumber \\
&& -\int_0^L dx \rho_s(x)\ln \left[4\pi\frac{\sinh p_s |\frac{\partial \psi^{\star}(x)}{\partial x}|}{p_s |\frac{\partial \psi^{\star}(x)}{\partial x}|}\right]
+ \sum_{j=\pm,s}\int_0^L dx \rho_j(x) \left[\ln \rho_{j}(x)-1\right] \label{eq:free_saddle_den1d}
\end{eqnarray}
and
\begin{eqnarray}
\frac{F_{chem}^{\star}}{Ak_B T} &=& -\sigma_1 \psi^{\star}(x_1)-\sigma_2 \psi^{\star}(x_2) \label{eq:saddle_chemical_free}
\end{eqnarray}
In order to compute the electrocaloric effect, free energy \textit{changes} with respect to the system in
the absence of applied electric field are desirable. In the absence of applied electric field
(i.e., when $\sigma_1 = \sigma_2=0$ and considered as the reference state), the \textit{same} number of ions
and solvent molecules are homogeneously distributed in
volume $\Omega = AL$ so that free energy of the reference state becomes
\begin{eqnarray}
\frac{F_{h}^{\star}}{AL k_B T} = \left[\frac{F^{\star} + F_{chem}^{\star}}{A L k_BT}\right]_{\sigma_1 = \sigma_2=0} 
&=& 2\rho_{c,b}\left[\ln \rho_{c,b}-1\right] + \rho_{s,b}\left[\ln \rho_{s,b}-1 - \ln 4\pi\right] \label{eq:free_homo}
\end{eqnarray}
where, we have used the constraint $A\int_0^L dx \rho_{j,x}= \rho_{j,b}\Omega$ for equating the number of ions and
solvent molecules in the absence and presence of applied electric field. Using these equations, the free energy
changes ($\Delta F^{\star}$) due to the application of an electric field can be written as
\begin{eqnarray}
\frac{\Delta F^{\star}}{A k_B T} = \frac{F_{dl}^{\star}-F_{h}^{\star}}{A k_B T} &=& \frac{F^{\star}-F_{h}^{\star}}{A k_B T} - \sigma_1 \psi^{\star}(x_1) - \sigma_2 \psi^{\star}(x_2) \label{eq:define_deltaF}
\end{eqnarray}

\subsection{Spatially uniform dielectric : Poisson-Boltzmann (PB) and modified Poisson-Boltzmann (MPB) approaches}
In the limits of small surface potentials so that $\psi^{\star}(x)-\psi_b^{\star} \rightarrow 0$ and
weak coupling limt for dipoles, defined by $p_s |\frac{\partial \psi^{\star}(x)}{\partial x}| \rightarrow 0$,
the dielectric function given by Eq. ~\ref{eq:dielec_function} becomes spatially uniform so that
\begin{eqnarray}
\epsilon(x)\equiv \epsilon_h &=& 1 + \frac{4\pi}{3} l_{Bo}p_s^2 \rho_{s,b}
\end{eqnarray}
Physically, this means that solvent density is spatially uniform in the limits of small surface potentials and weak coupling limit for dipoles so that $\rho_s(x) = \rho_{s,b}$ as evident from
Eq. ~\ref{eq:solv_den}. It is to be noted that the quantity $f$ is taken to be unity in these limits and leads to
the standard Poisson-Boltzmann results pioneered by Verwey and Overbeek\cite{verweybook}. Another somewhat recent
development (so called modified Poisson-Boltzmann (MPB) approach\cite{kornyshev}) is
to consider the case of uniform dielectric but \textit{include steric effects in the calculations of charge density}
by taking
\begin{eqnarray}
f\left\{\psi^{\star}(x)-\psi^{\star}_b,\frac{\partial \psi^{\star}(x)}{\partial x}\right\} &\equiv& 
f_{MPB}\left\{\psi^{\star}(x)-\psi^{\star}_b\right\} =  1-\alpha_0 + \alpha_0 \cosh\left[|Z_c|\left\{\psi^{\star}(x)-\psi^{\star}_b\right\}\right]\nonumber \\
&&
\end{eqnarray}, where $\alpha_0 = 2\rho_{c,b}a^3$ is the packing fraction of ions in the bulk. Although it seems inconsistent
to ignore and retain functional dependence of a particular quantity such as $f$ while considering different physical quantities such as dielectric function
and charge density, the MPB approach has been quite successful in predicting qualitative features of the
double layer capacitance. Nevertheless, the MPB approach leads to semi-analytical predictions for the electrostatic potential and the free energy, as described below.

With the approximations described above, Eq. ~\ref{eq:saddle2_1d} can be readily integrated over
$x$ (after multiplying by $\frac{\partial \psi^{\star}(x)}{\partial x}$ on both sides).
In particular, we obtain a self-consistent equation for $\frac{\partial \psi^{\star}(x)}{\partial x}$
\begin{eqnarray}
\frac{1}{2}\left[\frac{\partial \psi^{\star}(x)}{\partial x}\right]^2 &=& \frac{4\pi l_{Bo}}{\epsilon_h a^3} \left[\ln f_{MPB}\left\{\psi^{\star}(x)-\psi^{\star}_b\right\} -\lambda\right]\label{eq:first_integral}
\end{eqnarray}
where $\lambda$ is an integration constant, which is determined below and the effects of surface charge densities
($\sigma_1, \sigma_2$) appear in the form of boundary conditions.
Using Eq. ~\ref{eq:first_integral} and
equations at the saddle-point, the free energy changes of the double layer system, defined by
Eq. ~\ref{eq:define_deltaF}, can be written as
\begin{eqnarray}
\frac{\Delta F^{\star}_{MPB}}{Ak_B T} &=& \frac{F_{dl,MPB}^{\star}-F_{h}}{A k_B T} 
= -\frac{\lambda L}{a^3} - \frac{2}{a^3}\int_{\psi^{\star}(x_1)}^{\psi^{\star}(x_2)}d\psi \frac{\left[\ln f_{MPB}\left\{\psi^{\star}(x)-\psi^{\star}_b\right\} -\lambda\right]}{\frac{\partial \psi^{\star}(x)}{\partial x}} \label{eq:free_lambda}
\end{eqnarray}
where $F_{dl,MPB} = F^{\star}_{MPB} + F_{chem}^{\star}$ and $F^{\star}_{MPB}$ is the approximation for
Eq. ~\ref{eq:free_saddle_den1d} obtained using Eq. ~\ref{eq:first_integral} and $F_{chem}^{\star}$
is given by Eq. ~\ref{eq:saddle_chemical_free}. We must point out that
in obtaining Eq. ~\ref{eq:free_lambda}, we have retained functional dependence of the
solvent density on $f_{MPB}$ through Eq. ~\ref{eq:solv_den} and used the incompressibility constraint.

In the following, we consider two cases of non-overlapping and overlapping double layers and eliminate $\lambda$ from Eq.
\ref{eq:free_lambda}. In the case of non-overlapping double layers, $\psi^{\star}(x)$ becomes a non-monotonic
function of $x$ with a minimum at $x=x_{min}$. Integrating Eq. ~\ref{eq:first_integral} over $x$ with the limits $x_1$ and $x_2$, we obtain\cite{overbeekPaper}
\begin{eqnarray}
\int_{\psi^{\star}(x_{min})}^{\psi^{\star}(x_1)} \frac{d\psi^{\star}}{\left[\ln f_{MPB}\left\{\psi^{\star}(x)-\psi^{\star}_b\right\} -\lambda\right]^{1/2}} + \int_{\psi^{\star}(x_{min})}^{\psi^{\star}(x_1)} \frac{d\psi^{\star}}{\left[
\ln f_{MPB}\left\{\psi^{\star}(x)-\psi^{\star}_b\right\} -\lambda\right]^{1/2}} && \nonumber \\
= \left[\frac{8\pi l_{Bo}}{\epsilon_h a^3}\right]^{1/2} L \quad \quad \quad &&\label{eq:second_integral_nm}
\end{eqnarray}
Similarly, for the case of overlapping double layers so that $\psi^{\star}(x_1) > \psi^{\star}(x_2)$, we obtain
\begin{eqnarray}
\int_{\psi^{\star}(x_{2})}^{\psi^{\star}(x_1)} \frac{d\psi^{\star}}{\left[\ln f_{MPB}\left\{\psi^{\star}(x)-\psi^{\star}_b\right\} -\lambda\right]^{1/2}} &=& \left[\frac{8\pi l_{Bo}}{\epsilon_h a^3}\right]^{1/2} L \label{eq:second_integral_m}
\end{eqnarray}
Eqs. ~\ref{eq:second_integral_nm} and ~\ref{eq:second_integral_m} allows us to eliminate $\lambda$ from Eq. ~\ref{eq:free_lambda} and write it as
\begin{eqnarray}
\frac{\Delta F^{\star}_{MPB}}{A k_B T} &=& - \sqrt{\frac{\epsilon_h}{2\pi l_{Bo}a^3}}g\left\{\psi^{\star}(x_1),\psi^{\star}(x_2)\right\} = 
- \frac{4\rho_{c,b}|Z_c|}{\sqrt{2 \kappa^2 \alpha_o }}g\left\{\psi^{\star}(x_1),\psi^{\star}(x_2)\right\}\label{eq:free_final}
\end{eqnarray}
where we have defined $\kappa^2 = 8\pi l_{Bo} |Z_c|^2 \rho_{c,b}/\epsilon_h$. Also,
\begin{eqnarray}
g\left\{\psi^{\star}(x_1),\psi^{\star}(x_2)\right\} &=& \sum_{k=1,2}\int_{\psi^{\star}(x_{min})}^{\psi^{\star}(x_k)}d\psi \sqrt{\ln f_{MPB}\left\{\psi^{\star}(x)-\psi^{\star}_b\right\}}
\end{eqnarray}
for the non-overlapping double layers and
\begin{eqnarray}
g\left\{\psi^{\star}(x_1),\psi^{\star}(x_2)\right\} &=&
\int_{\psi^{\star}(x_{2})}^{\psi^{\star}(x_1)}d\psi \sqrt{\ln f_{MPB}\left\{\psi^{\star}(x)-\psi^{\star}_b\right\}}
\end{eqnarray}
in the case of overlapping double layers.

Changes in entropy ($\Delta S^{\star}_{MPB}$) can be readily calculated using Eq. ~\ref{eq:free_final} and the thermodynamic relation
$\Delta S = -\left(\frac{\partial \Delta F}{\partial T}\right)_{\Omega}$ so that
\begin{eqnarray}
\frac{\Delta S^{\star}_{MPB}}{A k_B} &=& -\left(\frac{\partial \Delta F^{\star}_{MPB}}{A k_B\partial T}\right)_{\Omega}
 = \frac{6\rho_{c,b}|Z_c|}{\sqrt{2 \kappa^2 \alpha_o }}g\left[1 + 
\frac{2T}{3 g}\frac{\partial g}{\partial T}\right]\label{eq:entropy_final}
\end{eqnarray}
where we have dropped explicit functional dependencies of $g$ on $\psi^{\star}(x)-\psi^{\star}_b$ for convenience in writing. It is interesting to consider the limit of dilute solutions so that $\alpha_0\rightarrow 0$ and this
limit is the same as the standard PB approach. In this limit,
for non-overlapping double layers, $\psi^\star(x_{min}) = \psi^{\star}_b$ and $\lambda = 1$ (due to the fact that
$\partial \psi^{\star}(x)/\partial x = 0$ at $x = x_{min}$ in Eq. ~\ref{eq:first_integral}). This leads to
\begin{eqnarray}
\frac{\Delta F^{\star}_{PB}}{A k_B T} &=& \frac{4\rho_{c,b}}{\kappa}\sum_{k=1,2}\left[2-2\cosh\left(|Z_c|\left\{\psi^{\star}(x_k)-\psi^{\star}_b\right\}\right)\right]\label{eq:free_final_PB}
\end{eqnarray}
i.e., the total free energy change is the sum of changes in the individual double layers\cite{overbeekPaper}.
This leads to entropic changes given by
\begin{eqnarray}
\frac{\Delta S^{\star}_{PB}}{A k_B} &=& -\frac{4\rho_{c,b}}{\kappa}\sum_{k=1,2}\left[3-2\cosh\left(\frac{|Z_c|}{2}\left\{\psi^{\star}(x_k)-\psi^{\star}_b\right\}\right) - \sech\left(\frac{|Z_c|}{2}\left\{\psi^{\star}(x_k)
-\psi^{\star}_b\right\}\right) \right . \nonumber \\
&& \left . 
+ \left\{1- \sech\left(\frac{|Z_c|}{2}\left\{\psi^{\star}(x_k)-\psi^{\star}_b\right\}\right)\right\}\frac{T}{\epsilon_h}
\frac{\partial \epsilon_h}{\partial T}\right]\label{eq:entropy_final_PB}
\end{eqnarray}

\section{Numerical methods}
We have solved the set of equations (Eqs. ~\ref{eq:saddle2_1d}-~\ref{eq:solv_den}) numerically after rewriting Eq. ~\ref{eq:saddle2_1d} in the form
\begin{eqnarray}
\frac{\partial \psi^{\star}(x)}{\partial t} &=& \frac{\partial^2 \psi^{\star}(x)}{\partial x^2} + \frac{1}{\epsilon(x)}\frac{\partial \epsilon(x)}{\partial x} \frac{\partial \psi^{\star}(x)}{\partial x} + \frac{4\pi l_{Bo}}{\epsilon(x)}\rho_e(x)\label{eq:PB_numerical}
\end{eqnarray}
where $t$ is a fictitious time. A steady state solution of Eq. ~\ref{eq:PB_numerical} is obtained by using the extrapolated gear\cite{kumar_kilbey}
scheme and using size of ions $a$ to obtain dimensionless length variables. Time step of $0.0001$ is used to integrate Eq. ~\ref{eq:PB_numerical} with $L/a = 20 - 40$ (depending on the value of $\rho_{c,b}$) and $1024$ grid points. Convergence
of the numerical solution is checked by computing free energy changes
between two consecutive time steps and the changes less than
$0.0001$ are used to set the tolerance criteria. These equations are solved for non-overlapping double layer systems so that
one of the surfaces has the known surface potential while the other is grounded (i.e., surface potential is zero).
The temperature is changed by varying $l_{Bo}$ and the free energy changes (in units of $Ak_B T/a^2$) are computed
using Eqs. ~\ref{eq:free_saddle_den1d}, ~\ref{eq:saddle_chemical_free},
~\ref{eq:free_homo} and ~\ref{eq:define_deltaF}. In computing the electrocaloric
effect, we have made use of the
fact that the field variable $\psi^{\star}(x)$ in the theory is the
electrostatic potential (in units of $k_B T/e$) at location $x$. For example,
$\psi^{\star}(0) = eV_{0,1}/k_B T$ for the single double layer system studied in this work.
Numerical estimates for the surface charge densities we obtained by
the relation $\sigma =-[\epsilon(x)/4\pi l_{Bo})(\partial \psi^{\star}(x)/\partial x]_{x=0}$.

\section{Results: anatomy of the double layer}
\begin{figure}[htbp]  \centering
\subfigure[]{\includegraphics[width=3in]{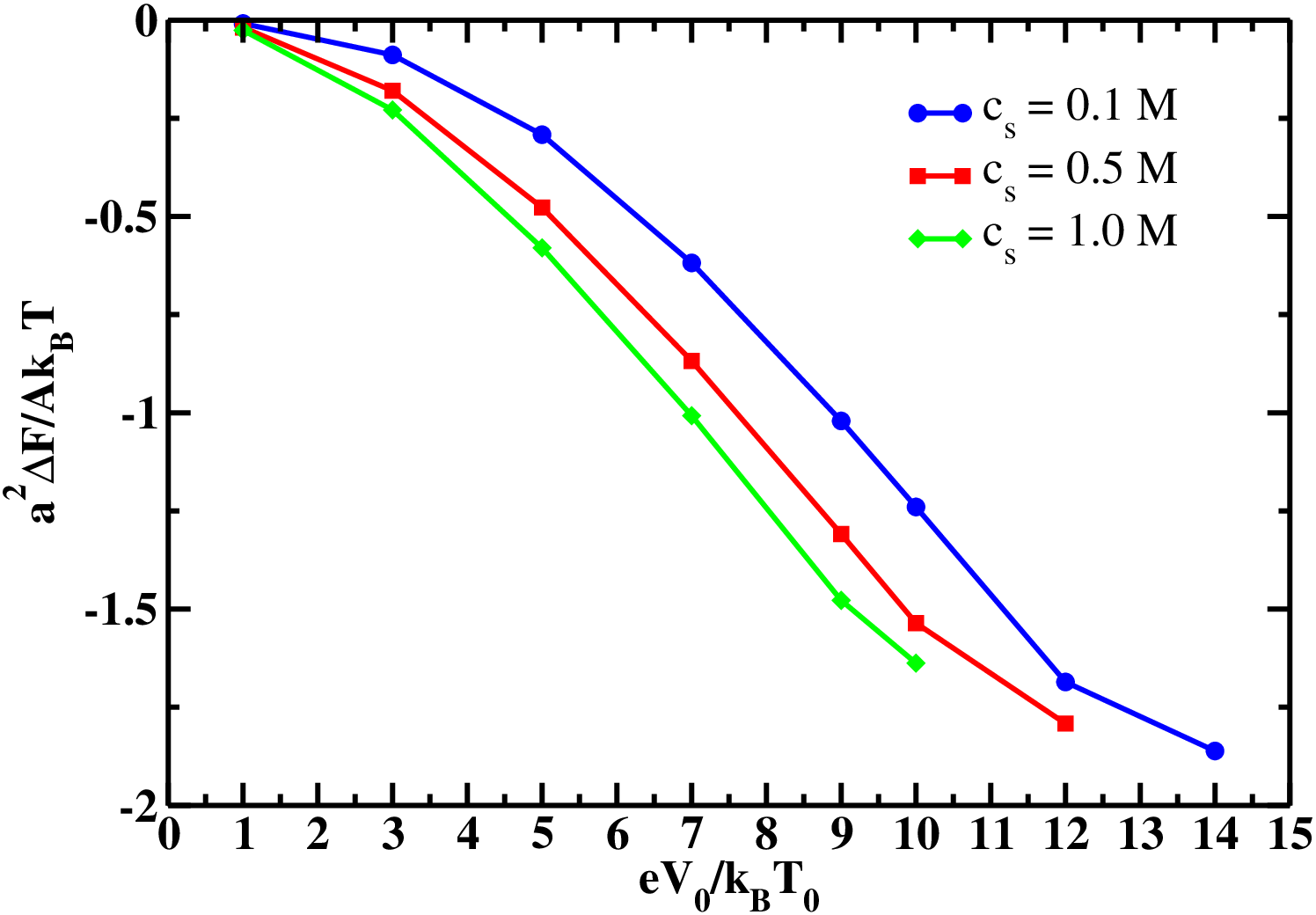}}\hspace{0.1in}
\subfigure[]{\includegraphics[width=3in]{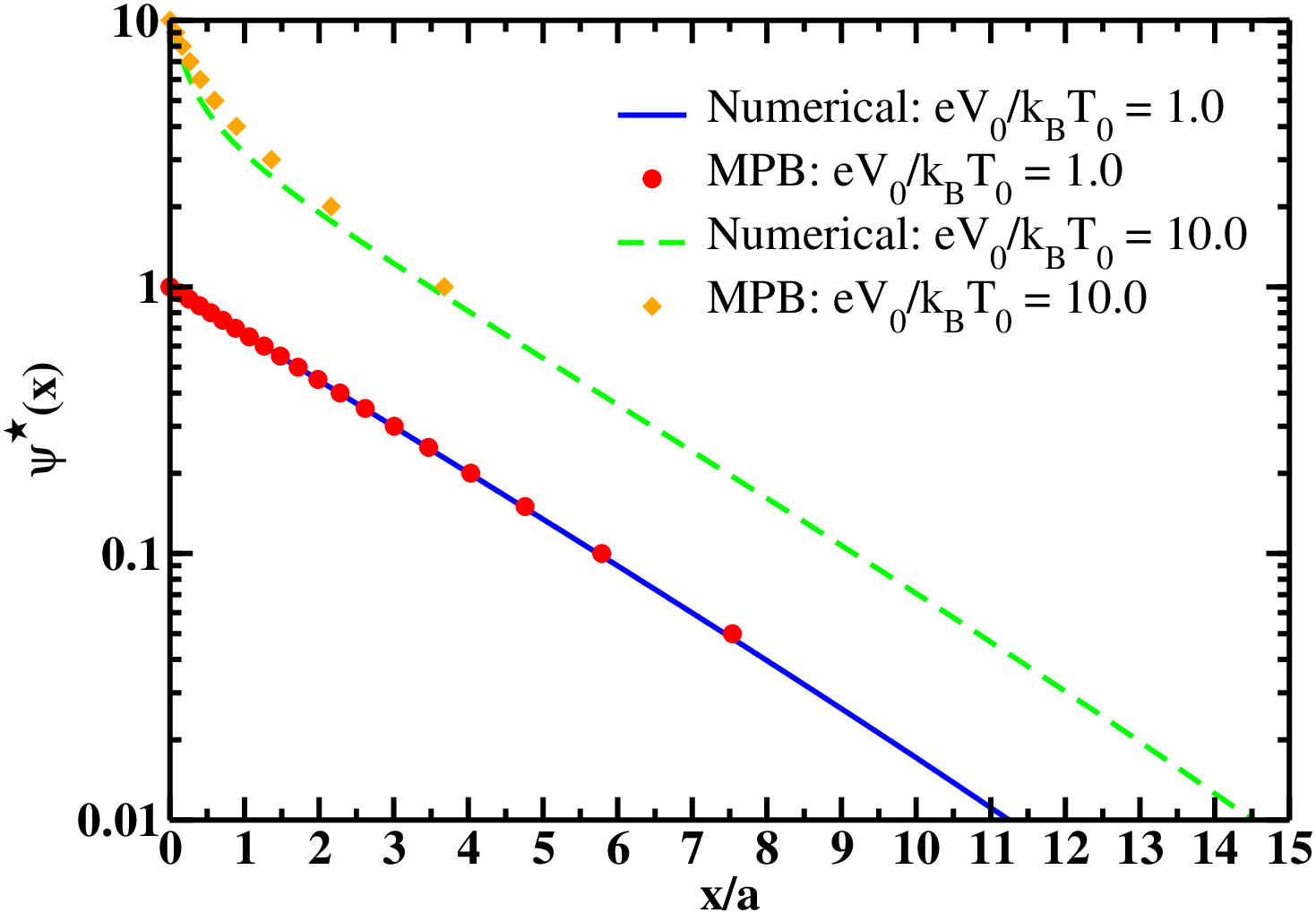}}
\caption{(a) Effects of the bulk salt concentration on the free energy changes ($\Delta F = \Delta F^{\star}$)
of the double layer at $T_0 = 303$ K.
(b) Comparisons of electrostatic potential profiles ($\psi^{\star}(x)$) from the MPB approach and numerical calculations at $c_s = 0.1$ M.}
\label{fig:free_energy_all}
\end{figure}

\begin{figure}[htbp]  \centering
\subfigure[]{\includegraphics[width=3in]{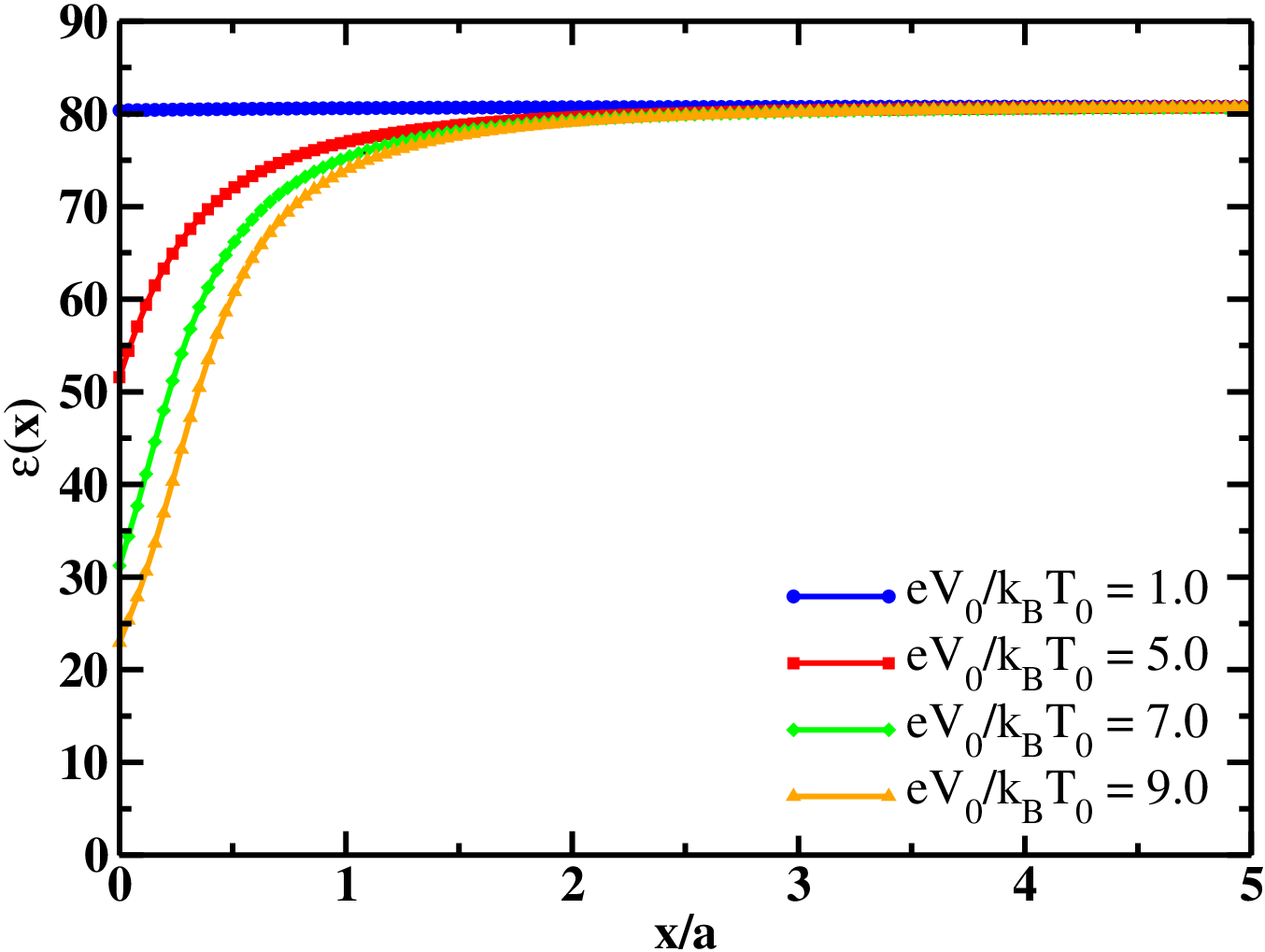}}\hspace{0.1in}
\subfigure[]{\includegraphics[width=3in]{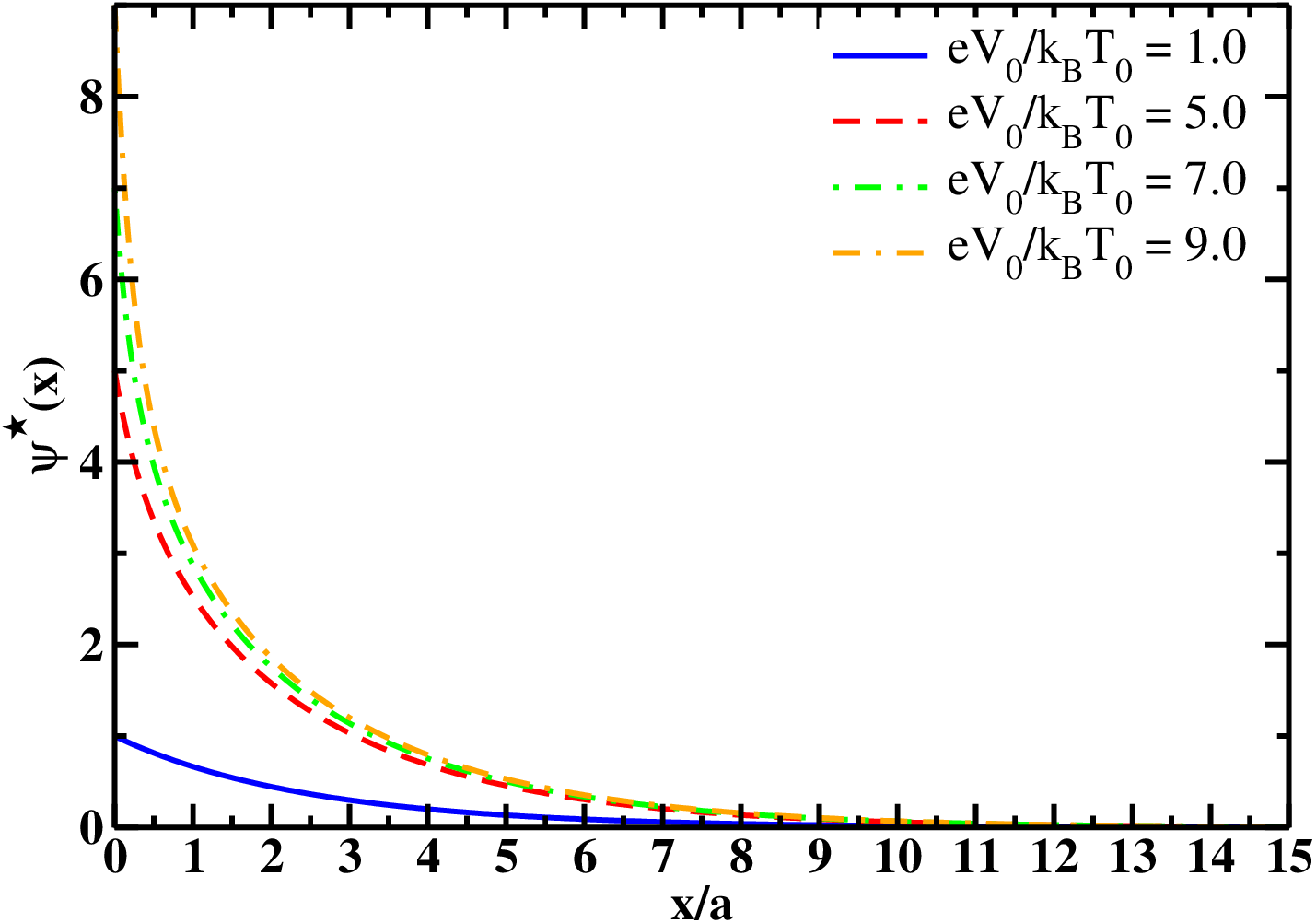}}\vspace{0.2in}\\
\subfigure[]{\includegraphics[width=3in]{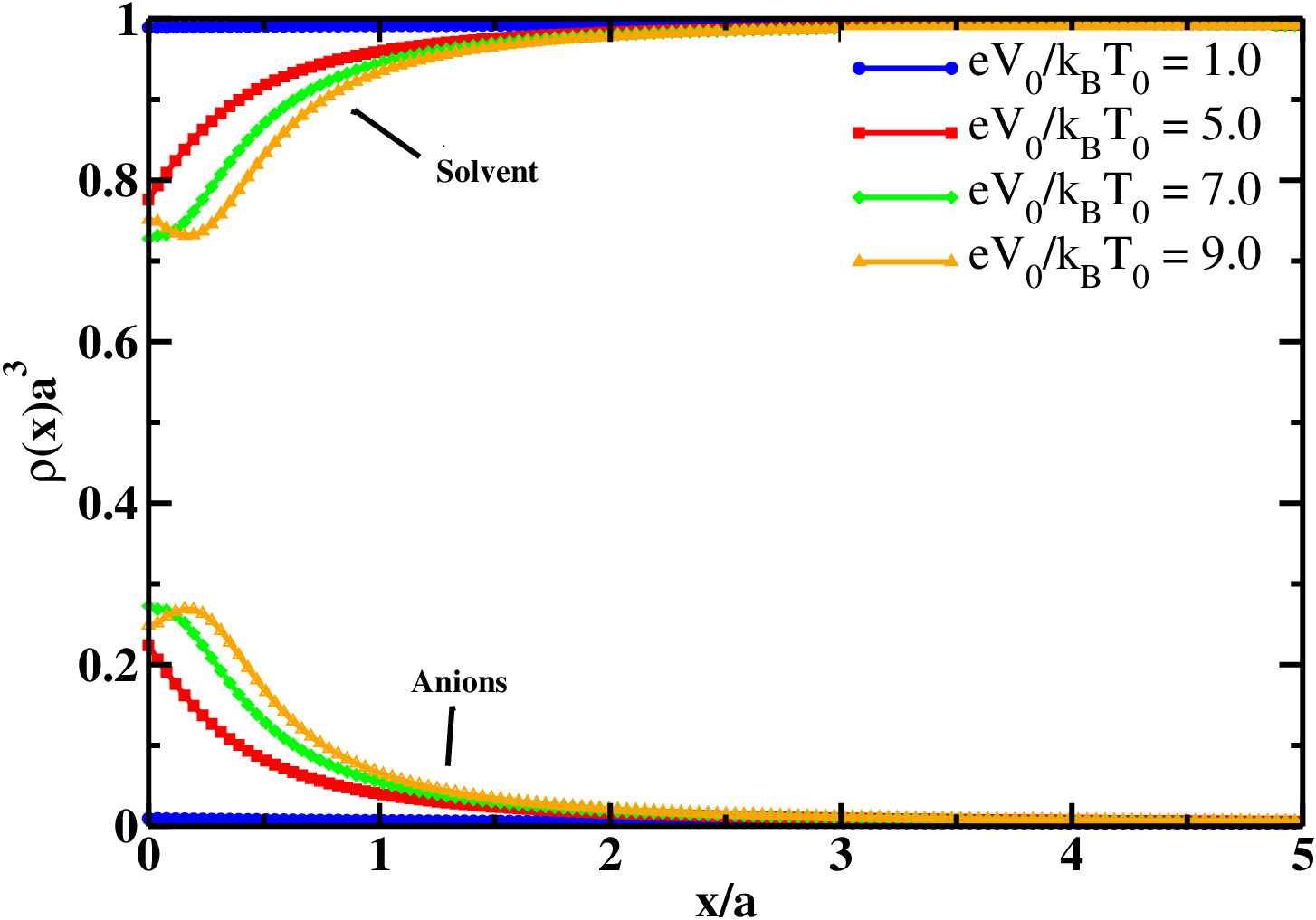}}\hspace{0.1in}
\subfigure[]{\includegraphics[width=3in]{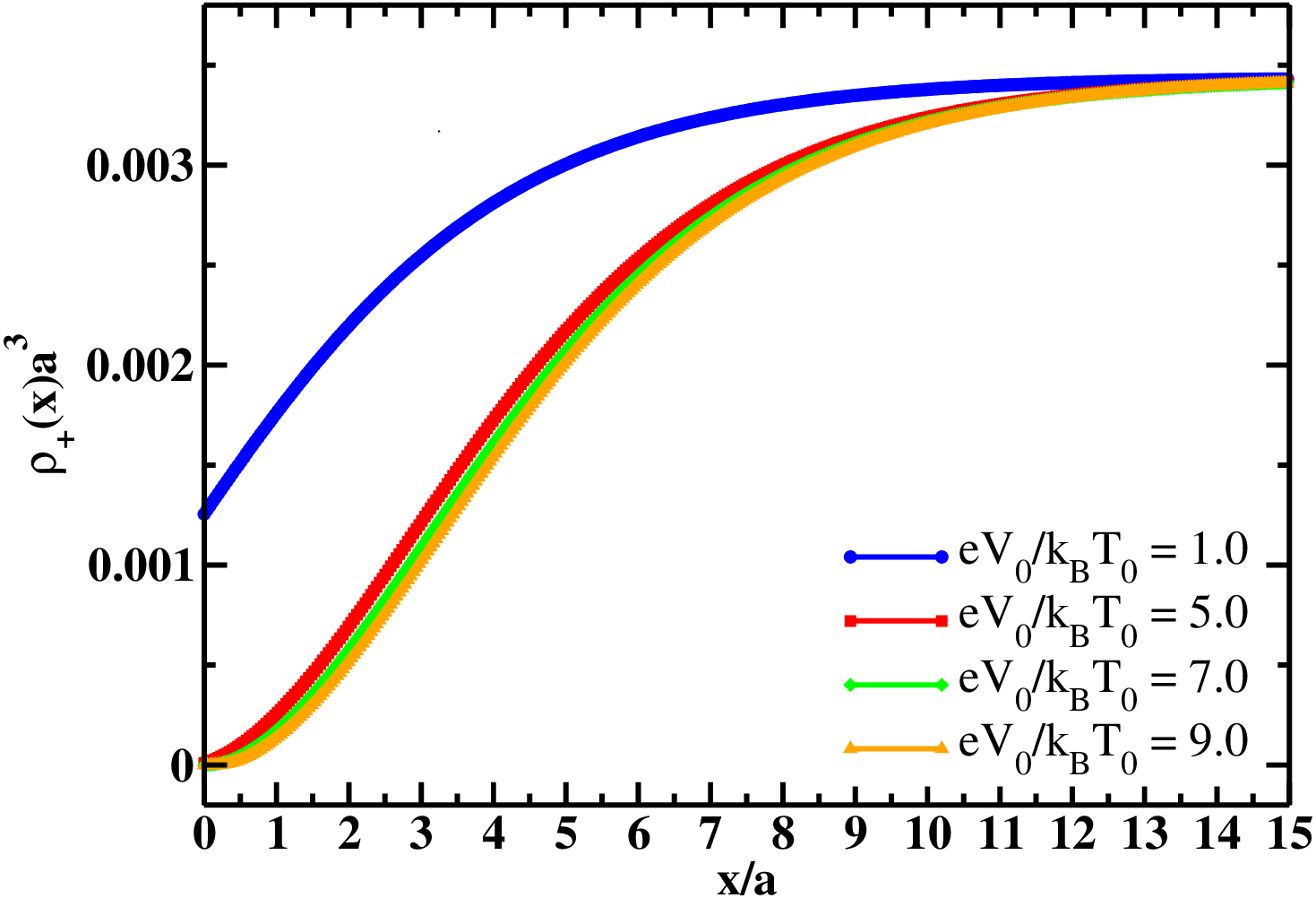}}
\caption{(a) The dielectric function, (b) electrostatic potential, (c) solvent and counterion (anion) densities, and (d)
co-ion (cation) densities at different surface potentials are shown for bulk salt concentration of $0.1$ M at $T_0 = 303$ K.}
\label{fig:double_layer_T303_p1}
\end{figure}

Anatomy of the double layer is determined by the electrostatic potential profile. As the comparisons between the
PB and MPB approaches are well-known\cite{kornyshev}, we only show comparisons between the MPB and our numerical calculations in
Figure ~\ref{fig:free_energy_all}(b) for low and high surface potentials. It is found that
the MPB and numerical results are in excellent agreement at $eV_0/k_B T_0 = 1$ showing exponential decay appearing as
linear on semi-log plot, as
expected. In contrast, the electrostatic potential profiles differ near the surface (for $x/a<2$) at $eV_0/k_B T_0 = 10$,
which are responsible for differences in free energies predicted using the MPB approach and the numerical calculations (cf.
Figure $3$(a) in the main text).

The differences in the electrostatic potential near the surface show up in plots for
surface charge density ($\sigma_1 = \sigma$) as a function of applied surface potential (Figure $3$(b) in the main text). The structural changes resulting from an increase in the surface potentials are shown in Figure ~\ref{fig:double_layer_T303_p1}. In particular, an increase in surface potentials leads to an increase in
the volume fraction of counterions (anions in this case) near the surface at the expense of excluding
coions and solvent molecules. However, further increase in the surface potential (e.g., see plot for $eV_0/k_B T_0 =9$ in
Figure ~\ref{fig:double_layer_T303_p1}(c)) leads to increase in solvent volume fraction near the surface at the expense of
exclusion of counterions and coions. This is expected from the expression for solvent volume fraction, Eq. ~\ref{eq:solv_den}, leading to higher volume fraction of solvent in regions having strong electric fields.
Also, such an enrichment of solvent in regions of strong electric fields is in agreement with
previous theoretical works\cite{electrosorption1,electrosorption2}.
Furthermore, the electric field dependent sorption of water on the AFM tips has been used to modulate friction at
the nanoscale\cite{evgheniFRIC}.

  \setcounter{equation}{0}  
\renewcommand{\theequation}{A-\arabic{equation}}
  \setcounter{equation}{0}  
  \section*{APPENDIX A : Field theory for double layer systems in the grand canonical ensemble} \label{app:A}
For study of a double layer, grand canonical partition function can be constructed and is given by
$\Gamma = \sum_{j=\pm,s}e^{\mu_j n_j/k_B T}Z\left\{n_j\right\}$ so that

\begin{eqnarray}
       \Gamma & = & \frac{1}{N_\psi}\int D\left[\psi\right]\int  D\left[\eta\right]
\exp \left [- \frac{H_g\left\{\psi,\eta\right\}}{k_BT}\right ] \label{eq:parti_grand}
\end{eqnarray}
so that
\begin{eqnarray}
       \frac{H_g\left\{\psi,\eta\right\}}{k_BT} &=& - \frac{1}{8\pi l_{Bo}}\int d\mathbf{r} \psi(\mathbf{r})\nabla_{\mathbf{r}}^2 \psi(\mathbf{r})
- i \int d\mathbf{r} \eta (\mathbf{r}) + \sigma_1 \int d\mathbf{r_{\parallel}}i \psi(\mathbf{r_{\parallel}},x_1) \nonumber \\
&& + \sigma_2 \int d\mathbf{r_{\parallel}}i \psi(\mathbf{r_{\parallel}},x_2)
- \sum_{j=\pm,s}e^{\mu_j/k_B T} Q_{j}\left\{\psi,\eta\right\} \label{eq:hami_grand}
\end{eqnarray}
where we have used Eqs. ~\ref{eq:parti_physical} and ~\ref{eq:hami_physical}
for the partition function in the canonical ensemble.

The saddle point approximation with respect to $\eta$ and $\psi$ gives two non-linear equations. At the saddle-points,
both $\eta$ and $\psi$ turn out to be purely imaginary. Writing $i\eta(\mathbf{r}) = \eta^\star(\mathbf{r})$
and $i\psi(\mathbf{r}) = \psi^\star(\mathbf{r})$ at the saddle point, the two equations are given by
\begin{eqnarray}
       \sum_{j=\pm}v_j \rho_j(\mathbf{r}) + v_s \rho_s(\mathbf{r}) &=& 1  \label{eq:saddle1_g}\\
\nabla_{\mathbf{r}}\cdot\left[\epsilon(\mathbf{r})\nabla_{\mathbf{r}}\psi^{\star}(\mathbf{r})\right] &=& 
-4\pi l_{Bo} \rho_e(\mathbf{r}) \label{eq:saddle2_g}
\end{eqnarray}
where we have defined
\begin{eqnarray}
       \rho_{j=\pm}(\mathbf{r}) &=& \exp \left [\frac{\mu_j}{k_B T} - Z_j \psi^{\star}(\mathbf{r}) - v_j\eta^\star(\mathbf{r})\right ] \label{eq:den_ions_gc}\\
 \rho_{s}(\mathbf{r}) &=& 4\pi \exp\left[\frac{\mu_s}{k_B T} - v_s \eta^\star(\mathbf{r})\right] \frac{\sinh p_s |\nabla_{\mathbf{r}}\psi^{\star}(\mathbf{r})|}{p_s |\nabla_{\mathbf{r}}\psi^{\star}(\mathbf{r})|} \label{eq:den_solvent_gc}
\end{eqnarray}
so that
$\rho_e(\mathbf{r}) = \sum_{j=\pm}Z_j \rho_{j}(\mathbf{r}) + \sigma_1 \delta(x-x_1) + \sigma_2 \delta(x-x_2)$ and
the local dielectric function is given by
\begin{eqnarray}
       \epsilon(\mathbf{r}) &=& 1 + 4\pi l_{Bo}p_s^2 \rho_s(\mathbf{r})
\frac{L\left[p_s |\nabla_{\mathbf{r}}\psi^{\star}(\mathbf{r})|\right]}{p_s |\nabla_{\mathbf{r}}\psi^{\star}(\mathbf{r})|}  
\end{eqnarray}
where $L(x) = \coth(x) - 1/x$ is the Langevin function. Corresponding approximation
for the Gibbs free energy is given by
\begin{eqnarray}
       \frac{H_g^{\star}}{k_BT} &=& \sigma_1 \int d\mathbf{r_{\parallel}} \psi^{\star}(\mathbf{r_{\parallel}},x_1) + \sigma_2 \int d\mathbf{r_{\parallel}}\psi^{\star}(\mathbf{r_{\parallel}},x_2) + \frac{1}{8\pi l_{Bo}}\int d\mathbf{r} \psi^{\star}(\mathbf{r})\nabla_{\mathbf{r}}^2 \psi^{\star}(\mathbf{r}) \nonumber \\
&& - \int d\mathbf{r} \eta^{\star} (\mathbf{r}) - \sum_{j=\pm,s}\int d\mathbf{r} \rho_j(\mathbf{r}) \label{eq:hami_grand_saddle}
\end{eqnarray}
Using Eqs. ~\ref{eq:saddle1_g},~\ref{eq:saddle2_g}, ~\ref{eq:den_ions_gc} and ~\ref{eq:den_solvent_gc}, it can be shown that $H_g^{\star}$ and $F^{\star}$ given by Eq. ~\ref{eq:free_saddle_den} are related by
\begin{eqnarray}
       \frac{F^{\star}}{k_BT} &=& \frac{H_g^{\star}}{k_BT} + \sum_{j=\pm,s}\frac{\mu_j}{k_B T}\int d\mathbf{r} \rho_j(\mathbf{r})
\end{eqnarray}
in accordance with the thermodynamic relation that the Helmholtz free energy is the Gibbs free energy plus chemical potential times the number of particles.

\section*{REFERENCES}
\setcounter {equation} {0}
\pagestyle{empty} \label{REFERENCES}

\end{document}